\documentclass[showpacs,preprint,prb,floatfix]{revtex4}
\usepackage{epsf}
\usepackage{bm}
\usepackage{amssymb}

\newcommand{\be}{\begin{equation}}
\newcommand{\ee}{\end{equation}}
\newcommand{\bea}{\begin{eqnarray}}
\newcommand{\eea}{\end{eqnarray}}
\newcommand{\beann}{\begin{eqnarray*}}
\newcommand{\eeann}{\end{eqnarray*}}
\newcommand{\bma}{\begin{array}{cc}}
\newcommand{\bmaccc}{\begin{array}{ccc}}
\newcommand{\ema}{\end{array}}
\newcommand{\fr}{\frac}
\newcommand{\ra}{\rangle}
\newcommand{\la}{\langle}
\newcommand{\li}{\left}
\newcommand{\re}{\right}

\newcommand{\alp}{\alpha}

\newcommand{\sig}{\sigma}
\newcommand{\gam}{\gamma}
\newcommand{\ny}{\nu}

\newcommand{\nn}{\nonumber}

\newcommand{\uarr}{\Uparrow}
\newcommand{\darr}{\Downarrow}
\newcommand{\suarr}{\uparrow} 
\newcommand{\sdarr}{\downarrow} 
\newcommand{\suarrbig}{\uparrow} 
\newcommand{\sdarrbig}{\downarrow} 

\newcommand{\ez}{\vec{e}_z}
\newcommand{\er}{\vec{e}_r}

\newcommand{\spinup}{|\!\uarr \ra}

\newcommand{\spindown}{|\!\darr \ra}

\newcommand{\sspinup}{|\suarrbig \ra}
\newcommand{\sspinupphi}{|\suarrbig(\varphi) \ra}
\newcommand{\sspindown}{|\sdarrbig \ra}
\newcommand{\sspindownphi}{|\sdarrbig(\varphi) \ra}

\newcommand{\rr}{r_0}   
\newcommand{\meff}{m^*}  
\newcommand{\epsc}{\varepsilon} 

\newcommand{\eftil}{\widetilde{E}_F}
\newcommand{\sqrteftil}{\widetilde{E}_F^{1/2}}
\newcommand{\mutil}{\widetilde{\mu}}
\newcommand{\keffrr}{k_{F} \rr}
\newcommand{\matt}{ {t} }
\newcommand{\mattt}{ {\tilde t} }
\newcommand{\matmd}{ {m}_3 }
\newcommand{\matNe}{ {N}_1 }
\newcommand{\matNz}{ {N}_2 }
\newcommand{\matS}{ {S} }
\newcommand{\matSre}{ {S}_{\rm lr} }
\newcommand{\matPi}{ P }

\newcommand{\mata}{ {P'}}

\begin{document}

 \title{Aharonov-Bohm Physics with Spin I: \\
 Geometric Phases in One-dimensional Ballistic Rings}

\author{Martina Hentschel}
\affiliation{
Duke University, Department of Physics, Box 90305, Durham, NC 27708-0305, U.S.A.}
\author{Henning Schomerus}
\affiliation{
Max-Planck-Institut f\"{u}r Physik komplexer Systeme, N\"othnitzer Stra\ss e 38,
01187~Dresden, Germany}
\author{Diego Frustaglia}
\altaffiliation[Present address: ]{NEST-INFM \& Scuola Normale Superiore, 56126 Pisa, Italy.}
\affiliation{Institut f\"ur Theoretische
Festk\"orperphysik, Universit\"at Karlsruhe, 76128 Karlsruhe, Germany}
\author{Klaus Richter}
\affiliation{ Institut f\"ur Theoretische Physik, Universit\"at
Regensburg, 93040 Regensburg, Germany}
\date{\today}
%
\begin{abstract}
{ We analytically calculate the spin-dependent electronic conductance through a
one-dimensional ballistic ring in the presence of an  inhomogeneous magnetic
field and identify signatures of geometric and Berry phases in the general
non-adiabatic situation. For an in-plane magnetic field, we rigorously prove the
spin-flip effect presented in Frustaglia {\it et al.}, Phys. Rev. Lett. {\bf
87}, 256602 (2001), which allows to control and switch the polarization of
outgoing electrons by means of an Aharonov-Bohm flux, and derive analytical
expressions for the energy-averaged magneto-conductance. Our results support
numerical calculations for two-dimensional ballistic rings presented in the
second paper (Frustaglia {\it et al.}, submitted to Phys. Rev. B) 
of this series.}
\end{abstract}

\maketitle

\bibliographystyle{simpl1}

%
\section{Introduction}

The Aharonov-Bohm (AB) effect\cite{AB59} represents a genuine interference
phenomenon at the heart of mesoscopic quantum physics.\cite{WW} It has allowed
to probe the coherence of wavefunctions extending over ring conductors of micron
scales by monitoring the magneto conductance as a function of a magnetic flux
threaded through the ring. To date the AB effect is being used as a tool to
investigate phase coherence and dephasing mechanisms in nanostructures. In
common AB setups composed of metal or semiconductor rings subject to uniform
flux-generating external magnetic fields, the relevant physics is governed by
(interference of) the orbital part of the electron wave function, while the spin
degree of freedom can usually be neglected. More recently, the role of the
electron spin as a means, beside charge, to control a current or to store
information has received much attention in the context of spintronics.\cite{P98}
Electrons with spin experience quantum phases beyond the charge-based AB phase.
The subject of this paper is the study of
the resulting complex interference phenomena on the theoretical side.



The adiabatic Berry phase \cite{berry_royalsoc} and non-adiabatic
Aharonov-Anandan geometric phases \cite{aharonov,shaperewilcek} are key aspects
of electronic transport in inhomogeneous magnetic fields. Just as other phases
(such as scattering phases in the Coulomb blockade regime
\cite{Yacoby:Heiblum:1995,Schuster:Buks:1997} and of Kondo
\cite{VanDerWiel:DeFranceschi:2000,Ji:Heiblum:2000,Ji:Heiblum:2002} or Fano
resonances \cite{Goeres:GoldhaberGordon:2000,Kobayashi:Aikawa:2002}) they can be
probed by interference experiments, which most easily are carried out in the AB
ring geometry. Since the spin degree of freedom becomes a dynamical quantity
in inhomogeneous magnetic fields, its generation of, 
and interplay with, geometric and Berry phases deserves detailed investigation.
\cite{LGB:1990,sternprl,sternlang,AL:1993,LAG:1998,diegoklaus,yang}

This is the first paper of a two-part series on spin-dependent electronic
transport through rings in the  presence of inhomogeneous magnetic fields.
In this first part we address ballistic spin-dependent transport
through circular rings in rotationally symmetric magnetic fields that occur
in realizations of inhomogeneous fields by a central micromagnet
\cite{ye} or by a current through a wire piercing the ring, \cite{current} see
Fig.~\ref{fig_magfeldgeom}. 
For narrow confinement the orbital transport channels decouple, and it
suffices to investigate the reduced strictly one-dimensional (1D) problem for
each channel, which, as we will show, can be solved exactly for all strengths
of the magnetic field.

The constraint of rotational invariance is relaxed
in the second part \cite{2dlang} of this series (also referred to as
paper II in the following), where three of us describe a general numerical Green
function approach to the spin-dependent magneto-conductance, which works for
arbitrary geometry of the system and texture of the magnetic field,  and also in
the presence of disorder. This approach is applied there to two-dimensional rings, 
and for a rotationally symmetric in-plane magnetic field a spin-flip effect is
found, which allows to control and switch the polarization of ballistic
electrons by a small Aharonov-Bohm flux through the ring (for a short exposition
of this effect see Ref.~\onlinecite{prl}).
In the present article (paper I) we provide a strict analytical proof of the
spin-flip effect, and derive analytical results that
support our findings for the two-dimensional (2D) rings presented in
paper II.


The outline of this paper is as follows. In Sec.~\ref{sec_1dring} we
introduce our model of a 1D ring subject to an
inhomogeneous magnetic field and present the
general solution of the Schr\"odinger equation. Section \ref{sec_magneto}
 is devoted
to the computation of the magneto-conductance within a transfer-matrix approach.
Generalizing the spin-independent transport discussed in
Sec.~\ref{subsec_spinindep}, spin-dependence is taken into account in
Sec.~\ref{subsec_spindep}. Results are presented in Sec.~\ref{sec_general},
where in
particular we consider a ring with a central micromagnet and discuss
the appearance of geometric phases in the magneto-conductance. Section
\ref{sec_inplane} deals with the special case of an in-plane
magnetic field. We analytically prove the spin-flip effect in
Sec.~\ref{subsec_spinswitch}. In Secs.~\ref{subsec_avgt} and
\ref{subsec_limiting} we discuss the transition from the non-adiabatic to the
adiabatic situation in terms of the energy-averaged conductance, and derive
simple analytical expressions that explain the observed features. We finish with
a discussion of our results in Sec.~\ref{sec_concl}.

\section {One-dimensional ring in an inhomogeneous magnetic field}
\label{sec_1dring}

We consider the spin-dependent
coherent electronic transport through a circular ring of radius $r_0$
within a layer of a two-dimensional electron gas (2DEG), exposed to
an inhomogeneous magnetic field $\vec{B}(\vec{r})$.
The transport is assumed to be ballistic,
i.e., the ring contains no impurities, and
electron-electron interactions are ignored.
The charge carriers in the 2DEG
are characterized by their electric charge $-e <0 $,
effective mass $\meff$, and magnetic moment $\mu = \fr{1}{2} \,g^* \,\mu_B$,
where
$g^*$ is the effective gyromagnetic ratio and $\mu_B = e \hbar / (2 m_0 c)$ is
Bohr's magneton with $m_0$ being the bare electron mass. For free
electrons in vacuum, the gyromagnetic ratio $g^* = g$ is approximately 2.
However, for electron-like quasi-particles in semiconductor heterostructures,
considerable deviations from this value occur depending on the material.
We set $\hbar=1$ throughout the rest of the paper and
introduce scaled parameters
for the magnetic moment,
$\widetilde {\mu} = 2 \meff \rr^2 \mu$, and
the Fermi energy in the ring,
$\widetilde{E}_F = 2 \meff \rr^2 \, E_F = (\keffrr)^2$.

The magnetic field $\vec{B}(\vec{r})$ couples to both the spin and the
orbital degrees of freedom. Spin-orbit interaction is assumed to be small, and will
be neglected. The effect of Rashba spin-orbit coupling \cite{lit_rashba}
is an additional in-plane magnetic field component
which depends on the Fermi energy. The
corresponding term is similar to the Aharonov-Casher
phase term considered in Ref.~\onlinecite{yi}, and not investigated here.
The  Hamiltonian within the confined region then reads
\be
\label{gen_hamiltonian}
H=\fr{1}{2 \meff}
       \li( \vec{p} + \fr{e}{c} \vec{A}_{\rm em}(\vec{r}) \re)^2
        + \mu \, \vec{\sigma} \cdot \vec{B}(\vec{r}) \:,
\ee
with $\vec{A}_{\rm em}(\vec{r})$ the vector potential, $\vec{B}(\vec{r}) =
\vec{\nabla} \times \vec{A}_{\rm em}(\vec{r})$, and $\vec{\sigma}$ the vector of
Pauli spin matrices. The first term in the Hamiltonian (\ref{gen_hamiltonian})
describes the kinetic energy and involves the generalized momentum $\vec{\Pi} =
\vec{p} + \fr{e}{c} \vec{A}_{\rm em}(\vec{r})$.
The second term $\mu\,\vec{\sigma} \cdot \vec{B}(\vec{r})$
corresponds to the Zeeman coupling of the
electron spin $\vec{\sigma}$ to the magnetic field $\vec{B}(\vec{r})$.

\begin{figure}[t]
  \epsfxsize=8.5cm
  \centerline{\epsffile{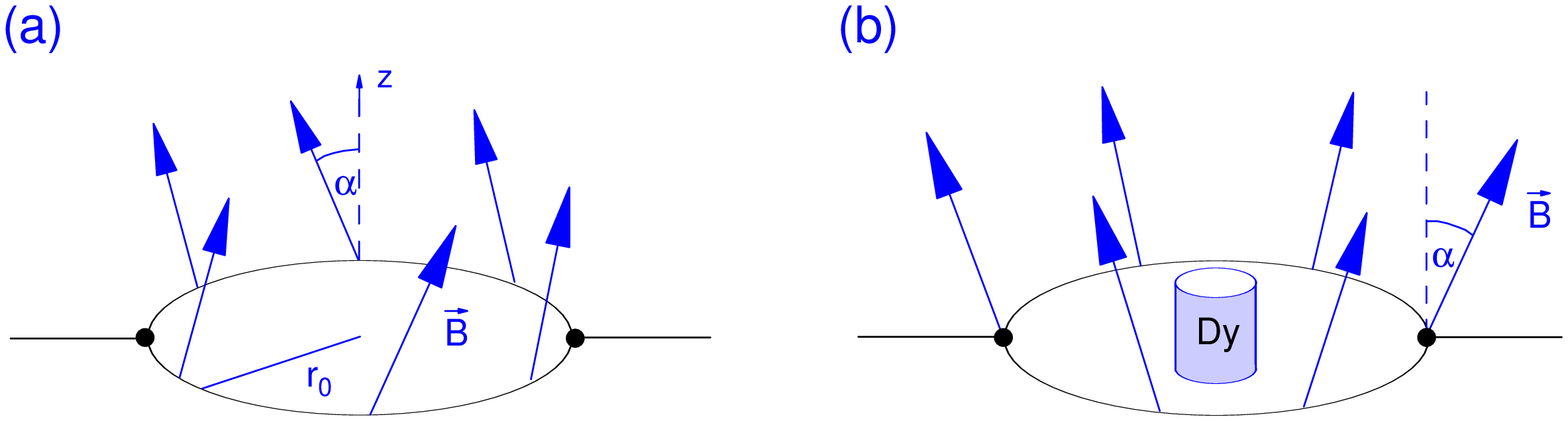}}
  \begin{center}
  \caption{Magnetic field texture for (a) a wire-like and (b) a crown-like
  magnetic field. The angle $\alp$ is defined
  as the angle of the magnetic field with respect to the $z$-axis.
  A possible way for creating a wire-like magnetic
  field in experiments is by means of a central current lead,
  while a crown-like field can be obtained by placing a micromagnet
  (like Dysprosium) into the center of the ring. Besides the homogeneous
  magnetic field in $z$-direction, a circular (tangential) or a radial magnetic
  component is then present, respectively.
  Both field textures are rotationally invariant about the $z$-axis.
  \label{fig_magfeldgeom}
  }  \end{center}
\end{figure}

We place the ring into the $xy$ plane and
decompose the magnetic field in polar coordinates
$\vec{B}=B_r \vec{e}_r+B_{\varphi} \vec{e}_\varphi+B_z \vec{e}_z$,
where the polar angle $\varphi$ parameterizes the position along
the ring. We
investigate the two magnetic field textures shown in
Fig.~\ref{fig_magfeldgeom}, which will be distinguished by
the parameter
\be\label{defphi0}
\varphi_t = \left\{\begin{array}{ll}
      \fr{\pi}{2} & \textrm {for a wire-like magnetic field,}\\
                0 & \textrm {for a crown-like magnetic field}.
                \end{array} \right.
\ee
The textures are further characterized by the tilt angle $\alpha$ of
the magnetic field
with respect to the perpendicular $z$-axis.
The total magnetic flux through the ring is denoted by
$\phi^{\rm AB}=\phi^{\rm AB}_z+\phi^{\rm AB}_{\rm ext}$, which includes the
contribution $\phi^{\rm AB}_z=\pi r_0^2 B_z$ generated by the
magnetic field in the ring itself,
as well as an extra magnetic flux $\phi^{\rm AB}_{\rm ext}$
that can be generated by an homogeneous magnetic field perpendicular
to the ring plane, or a solenoid piercing the ring.
The magnetic flux quantum is denoted by $\phi_0=e/2\pi$,
and the total strength of the magnetic field by $B=|\vec B|$.

The eigenstates
$|\Psi_n^{\suarr,\sdarr} \ra$
of the Hamiltonian (\ref{gen_hamiltonian}) can be
found analytically following Refs.~\onlinecite{sternprl,yi}.
They can be decomposed into an orbital part $\psi_n^{\suarr,\sdarr} (\varphi)$
and a spin part $|s_n^{\suarr,\sdarr} (\varphi)\ra$,
i.e., $\la\varphi|\Psi_n^{\suarr,\sdarr}\ra = \psi_n^{\suarr,\sdarr} (\varphi)
|s_n^{\suarr,\sdarr} (\varphi)\ra$.

The orbital functions $\psi_n^{\suarr,\sdarr}(\varphi)$ of the Hamiltonian
(\ref{gen_hamiltonian}) are plane waves $e^{i n \varphi}$.
Since later we will consider the effect of the Aharonov-Bohm flux $\phi^{\rm AB}$
and the transport through open rings, we will not require
the orbital quantum number $n$ to be an
integer, which equips us with a basis for scattering states
at arbitrary energy. This still leaves us with the problem
to find for a given energy
the correct values of $n$, as well as the spin directions.

The spin part
will be written in the $S_\alpha$-basis of spin
eigenstates $|\suarrbig(\varphi)\ra$, $|\sdarrbig(\varphi)\ra$ of the Zeeman term,
which in
turn read ($\varphi'\equiv \varphi+\varphi_t$)
\bea
{\langle \uarr |\suarrbig(\varphi)\ra \choose \langle \darr |\suarrbig(\varphi)\ra}
&=&
{\cos\fr{\alp}{2}\choose \:e^{ i \varphi'} \: \sin\fr{\alp}{2} },
\nonumber\\
{ \langle \uarr|\sdarrbig(\varphi)\ra \choose   \langle \darr|\sdarrbig(\varphi)\ra }
&=& {\sin\fr{\alp}{2}\choose -\:e^{ i \varphi'} \:
\cos\fr{\alp}{2} }\:
\label{spineigalp_unif}
\eea
in the $S_z$-basis $|\uarr\ra, |\darr\ra$ of the spin-up
and spin-down eigenstates of $\sig_z$.
The total eigenstates of the Zeeman term (including the orbital
part) will be denoted by $||\suarrbig_n\ra\ra$ and $||\sdarrbig_n\ra\ra$, with
$\la\varphi||\suarrbig_n\ra\ra=\exp(i n\phi)|\suarrbig(\varphi)\ra$
and $\la\varphi||\sdarrbig_n\ra\ra=\exp(i n\phi)|\sdarrbig(\varphi)\ra$.

The spin quantum number $\sigma=\suarr,\sdarr$ of
$|\Psi_n^{\sigma} \ra$ indicates whether the state
aligns parallel ($\sigma=\suarr$) or antiparallel ($\sigma=\sdarr$)
with the magnetic field when its strength is increased.
One then enters the adiabatic regime, in which the eigenstates
$|\Psi_n^{\sigma} \ra\sim ||\sigma_n\ra\ra$,
and the spins never
switch their direction with respect to the magnetic field
during the transport through the ring.
This requires a large ratio
\be\label{defQ}
Q = \fr{\omega_L}{\omega_{\rm orb}} = \fr{ {\widetilde {\mu}} B}{\keffrr}
\ee
of the spin precession (Larmor) frequency
\be\label{spinfreq}
\omega_L = \fr{ g^* e B}{ 2 m_0 }= \fr{\widetilde {\mu}B}{ \rr^2 \meff}
\ee
and the orbital frequency,
\be\label{orbfreq}
\omega_{\rm orb} = \fr{v_F}{ \rr}
= \fr{\keffrr}{\rr^2 \meff} \:.
\ee

In the adiabatic regime the two subspaces spanned by the states
$ || \suarrbig_n \ra\ra $ and
$ || \sdarrbig_n \ra\ra $
are completely decoupled.
In order to solve the problem under non-adiabatic conditions,
it is useful to
decompose the Hamiltonian (\ref{gen_hamiltonian}) into two parts, $H=H_0+H_1$,
where the adiabatic part $H_0$ contains no transitions between the
$||\suarrbig_n\ra\ra$ and $||\sdarrbig_n\ra\ra$ subspaces, whereas the
non-adiabatic part $H_1$ exclusively
describes such transitions.
To perform this decomposition, one can apply
the concept of a geometric vector potential $A_g$ introduced by Aharonov
and Anandan in Ref.~\onlinecite{aharonov}. The result
reads \cite{sternprl,sternlang}
\bea
H_0 & = &
\fr{1}{2 \meff} \li[ (\Pi - A_g)^2 + A_g^2 \re] + \mu \vec{\sig} \cdot \vec{B}
\nn \:, \\
H_1 & = &
\fr{1}{2 \meff} \li[(\Pi - A_g) A_g + A_g (\Pi - A_g) \re]
\label{H_0plusH_1}\:,
\eea
with the generalized momentum operator
\be
\Pi =  - \fr{i}{\rr} \fr{d}{d \varphi} + e A_{\rm em}^{\varphi}  
\ee
and the geometric vector potential
\be
\label{ag_mat}
A_g = \fr{\sin\alp}{2\rr}
\li( \bma
-\sin\alp & \cos\alp \: e^{-i \varphi'} \\
 \cos\alp \: e^{i \varphi'} & \sin\alp
\ema \re) \:.
\ee
The geometric vector potential $A_g$ causes the non-adiabatic
geometric \cite{aharonov,shaperewilcek} and adiabatic Berry phases. \cite{berry_royalsoc}
Note that only the direction $\alpha$ of the magnetic field at the position of the
ring enters the expression for $A_g$ explicitely, and that $\alpha$ changes under
variation of an external homogeneous magnetic field.
However, $A_g$ is not affected by solenoid-generated external Aharonov-Bohm fluxes
[that, of course, contribute to $\phi^{\rm AB}_{\rm ext}$ and $\phi^{\rm AB}$ and alter
the problem via, e.g., Eq.(\ref{nprn})].

The exact eigenstates of the
Hamiltonian (\ref{gen_hamiltonian})
now can be found in  the $S_\alpha$-basis
(\ref{spineigalp_unif}),
\[
\la \varphi|\Psi_n^{\suarr,\sdarr}\ra =
\psi_n(\varphi)
 \: \big[ C_{1,n}^{\suarr,\sdarr} \sspinupphi + C_{2,n}^{\suarr,\sdarr} \sspindownphi \big]
\:,
\]
where the coefficients $C_{1,n}^{\suarr,\sdarr}, C_{2,n}^{\suarr,\sdarr}$ are obtained from
the eigenvalue equation
\be\label{ewproblem}
\li( \bma
H_0^{\suarr \suarr} & H_1^{\suarr \sdarr} \\
H_1^{\sdarr \suarr} & H_0^{\sdarr \sdarr}
\ema \re)
  { C_{1,n}^{\suarr,\sdarr} \choose C_{2,n}^{\suarr,\sdarr} } \psi_n(\varphi)
= E_n   { C_{1,n}^{\suarr,\sdarr} \choose C_{2,n}^{\suarr,\sdarr} } \psi_n(\varphi) \:
\ee
with  $H_l^{\sig \sig'} = \la \sig(\varphi) | H_l | \sig'(\varphi) \ra$.
The diagonal elements describe the adiabatic part
of the problem,
whereas the non-diagonal entries contain the
non-adiabatic (spin-flip) processes that vanish in the adiabatic regime.

\begin{figure}[!t]
  \epsfxsize=5cm
  \centerline{\epsffile{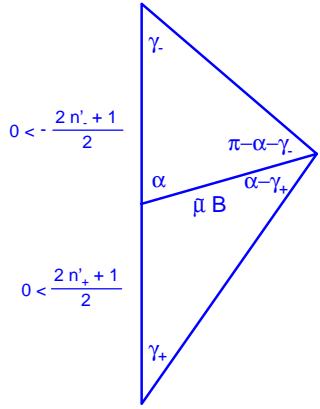}}
  \begin{center}
  \caption{Geometric interpretation of the angles $\gamma$ depending
  on the ratio of kinetic ($n'+1/2$) and magnetic ($\mutil B$)
  energy, and the (fixed)
  angle $\alp$. For $\alp > \pi/2$, the definitions of $n'_+$ and $n'_-$ have to be
  interchanged.
  \label{fig_dreiecknprmub}
  }  \end{center}
\end{figure}

Straightforward algebra yields that
${C_{1,n}^{\suarr} \choose C_{2,n}^{\suarr}}$ and
${C_{1,n}^{\sdarr} \choose C_{2,n}^{\sdarr}}$ are
the two eigenvectors of the matrix \cite{sternprl,sternlang}
\[
\li( \bma
n'^2+ \eta \sin^2 \fr{\alp}{2} + \widetilde {\mu} B & -\fr{\eta}{2}\sin \alp\\
-\fr{\eta}{2}\sin \alp
& n'^2+ \eta \cos^2 \fr{\alp}{2} - \widetilde {\mu} B
\ema \re)\:
\]
with $\eta = 2 n' +1$
and
\be\label{nprn}
n' = n + \phi^{\rm AB}/\phi_0\:,
\ee
which can be identified as the quantum number of the generalized momentum $\Pi$.
For given Fermi energy $\widetilde{E}_F$,
the four  solutions $n_{\rho}^{\prime\, \sigma}$ of the equation
\bea
0 & = & n'^4 + 2 n'^3 + ( 1 - 2 \eftil) n'^2 -
        2 (\eftil - \mutil B \cos \alp) n' \nn \\
  &   &   + \eftil^2 - \eftil + \mutil B \cos \alp -
        (\mutil B)^2 \label{polynomnpr}
\eea
can be associated to the four combinations
of counter-clockwise ($\rho=+$) and
clockwise ($\rho=-$) motion with spin up ($\sigma=\suarr$) or down  ($\sigma=\sdarr$).
The propagation sense is distinguished according to
the criterion $n^{\prime\sigma}_{+}+1/2 > 0$, $n^{\prime\sigma}_{-}+1/2 < 0$.
In the $S_z$-basis, the corresponding spin eigenstates are given by
\bea
\left( \begin{array}{c}
    \langle\uarr |s^\suarr_{n_\rho} \ra \\ \langle\darr |s^\suarr_{n_\rho} \ra \end{array}\right)
&=&
\li( \begin{array}{c}
                         \cos \fr{\gamma_\rho^{\suarr}}{2}
                         \\
                          \rho e^{i \varphi'} \sin
                         \fr{\gamma_\rho^{\suarr}}{2}
                         \end{array}
                         \re)
   , \nonumber\\
\left( \begin{array}{c}
    \langle\uarr |s^\sdarr_{n_\rho} \ra \\ \langle\darr |s^\sdarr_{n_\rho} \ra \end{array}\right)
&=&
 \li( \begin{array}{c}
                         \sin \fr{\gamma_\rho^{\sdarr}}{2}
                           \\
                         -\rho e^{i \varphi'} \cos
                         \fr{\gamma_\rho^{\sdarr}}{2}
                         \end{array}
                         \re)
,
  \label{basisstates}
\eea
with the angles
\be
{\rm cot}\gam_{\rho}^{\sigma}  =
\rho \li[ {\rm cot}\alp + \fr{(2{n'}_{\rho}^{\sigma}+1)}{2} \fr{1}{\widetilde {\mu} B \sin\alp}
\re]
\:.
\label{gamma1}
\ee
The spin states (\ref{basisstates}) are
of the form of Eq.~(\ref{spineigalp_unif}) with
$\alp$ replaced by $\gam_{\rho}^{\sigma}$. These angles
have an illustrative geometric interpretation shown in
Fig.~\ref{fig_dreiecknprmub}.
In the adiabatic limit, $\mutil B \gg |n'|$, we find
$\gamma_+^\sigma = \alp$ and $\gamma_-^\sigma = \pi - \alp$,
whereas in the diabatic limit $\gamma_\rho^\sigma = 0$.

This substitution rule $\alp \rightarrow \gamma_\rho^\sigma$ also applies to the
generalization of the adiabatic Berry phase  \cite{berry_royalsoc} to
non-adiabatic geometric Aharonov-Anandan phases: \cite{aharonov,shaperewilcek}
Here, $\gamma_\rho^\sigma$ replaces $\alp$ as the solid angle enclosed during
one round trip in parameter space. For the states $|\Psi_{n_+}^{\sigma}\ra$,
where $0 < \gamma_+^{\sigma} < \alp$, this nicely follows the intuitive
interpretation that in the non-adiabatic case the magnetic field is not strong
enough to force the spin into the direction of $\alp$, but only to a smaller
angle $\gamma_+^\sigma$. In this sense $\li | \fr{\gamma_+^\sigma-\alp}{\alp}
\re|$ can be taken as a qualitative measure for the deviation from the adiabatic
case: Starting with $\gamma_+^\sigma = \alp$ in the adiabatic situation, this
quantity decreases to eventually reach $\gamma_+^\sigma = 0$ in the diabatic
limit. Indeed, for the special case of an in-plane magnetic field
(Sec.~\ref{sec_inplane}) we will see that the adiabaticity parameter $Q$ defined
in Eq.~(\ref{defQ}) is directly related to the values of $\gamma_+^\sigma$,
cf.~Eq.~(\ref{gammaquerdeltan}).

\section{Magneto-conductance}
\label{sec_magneto}

\subsection{Spin-independent transport}
\label{subsec_spinindep}

We now investigate the magneto-conductance through a 1D ring within a transfer
matrix approach. For phase-coherent transport, the
conductance $G=(e^2/h) \sum_{\sigma,\sigma'} T_{\sigma \sigma'}$ is
given by the transmission probabilities $T_{\sigma \sigma'}$ between the spin channels
$\sigma,\sigma' = \suarr,\sdarr$ in each lead. We will work with the dimensionless conductance
$T= \sum_{\sigma,\sigma'} T_{\sigma \sigma'}$. In the adiabatic case, the transport is
spin-independent, and it suffices to study one spin species. \cite{imry} We
first review this case and then generalize it to the non-adiabatic situation,
partially following Ref.~\onlinecite{yi}.

We describe each of the two identical junctions ($i=1,2$) between
the leads and the 1D ring by a
$3\times3$ scattering matrix $\matS$, which relates,
by way of
\be\label{alpprSalp}
{\bf c}^{\prime (i)} = \matS {\bf c}^{(i)}
,
\ee
the coefficients
${\bf c}^{(i)}=(c^{(i)}_0,c^{(i)}_1,c^{(i)}_2)$
of incoming scattering states
to the coefficients
${\bf c}^{\prime (i)}=(c^{\prime (i)}_0,c^{\prime (i)}_1,c^{\prime (i)}_2)$
of outgoing scattering states
(cf.~Fig.~\ref{fig_yiamplitudes}).
Here the index $0$ is assigned to the coefficients in the  external leads,
the index $1$ denotes the upper arm of the ring, and the index $2$ denotes the lower arm
of the ring.
The scattering states are assumed to be
orthogonal
and normalized to carry unit particle flux.
Current conservation and time-reversal symmetry
at each junction imply that $\matS$
is unitary and symmetric,
and spatial symmetry leaves only one free parameter $\epsc$
(up to phase factors)
to characterize the coupling strength between the leads and the ring:
A wave is transmitted from the external leads into each of the two branches
of the ring with equal probability $\epsc$, whereas
reflection occurs with probability $1 - 2 \epsc$.
In particular, for $\epsc = 0$,
all particles are reflected so that there is no coupling into the
(then isolated) ring.

We write $\matS$ in the conventional form \cite{imry}
\be\label{smatrix}
\matS = \li( \bmaccc
                -(a+b) & \sqrt{\epsc} & \sqrt{\epsc}\\
                \sqrt{\epsc} & a & b \\
                \sqrt{\epsc} & b & a
              \ema
        \re)
\ee
where
\bea
a & = & \fr{1}{2} \li( \sqrt{ 1 - 2\epsc} -1 \re)\:,\nn \\
b & = &  a+1 \:. \nn
\eea

Before we turn to the case with spin where the amplitudes
${\bf c}^{(i)},{\bf c}^{\prime (i)}$ have spinor character,
let us summarize the result for the case
without spin from Ref.~\onlinecite{imry}.
In this case we can work with the orbital part alone, and the propagation velocities
in all scattering states at a given energy
are equal because there is no Zeeman energy, which simplifies the normalization
to unit particle flux.
Assuming $c_0^{(1)}=1, c_0^{(2)}=0$,
the dimensionless conductance $T=|t|^2$ is obtained from
\be
t = c_0^{\prime(2)} = - \fr{\epsc}{b^2} (1,1) \,\matt_1 \, \matPi {1 \choose -1}
\:. \label{tfromalp2}
\ee
with
\bea
\matPi & = & \li( \matSre \: \matt_2 \: \matSre \: \matt_1 - \openone_{2 \times 2}
        \re)^{-1} \:, \nn \\
\matSre & = & \fr{1}{b}\li( \bma a+b & a \\
                                -a & 1
                        \ema \re) \:. \nn
\eea
Here, $\matSre$ relates the amplitudes in the two arms of the ring
across the junctions,
whereas the transfer matrices $\matt_1,\matt_2$ relate the amplitudes within each
arm of the ring (see Fig.~\ref{fig_yiamplitudes}),
\bea
{c_2^{\prime(2)} \choose c_2^{(2)}}
 & =& \matSre {c_1^{(2)} \choose c_1^{\prime(2)}} \:, \: \nn
 \\
{ c_1^{(2)} \choose  c_1^{\prime(2)}}
& = &
\matt_1 { c_1^{\prime(1)} \choose c_1^{(1)}} \:, \quad
{c_2^{(1)} \choose c_2^{\prime(1)}} =
\matt_2 {c_2^{\prime(2)} \choose c_2^{(2)}}
\label{betasgammastransfer}\:.
\eea

For ballistic transport and symmetric arms,
the transfer matrices
\be
\label{thetadab}
\matt_1 = \matt_2 = e^{- i \theta_{\rm AB} }
                \li( \bma e^{i \theta_d} & 0 \\
                                0 & e^{- i \theta_d}
                        \ema \re) \:.
\ee
comprise two phase factors each,
namely the dynamic phase $\theta_d = \pi \keffrr$,
and an Aharonov-Bohm phase
$\theta_{\rm AB} = \pi \phi^{\rm AB} / \phi_0$
arising from the magnetic flux  through the ring.
For fixed Fermi energy, the dimensionless conductance  $T(\phi^{\rm AB})$ shows
characteristic Aharonov-Bohm fluctuations. \cite{imry}
The energy-averaged dimensionless conductance  ($\phi^{\rm AB}=0$) depends on the
coupling parameter $\epsc$ as
\be\label{avgtvonepsana}
\la T (\epsc) \ra = \fr{\epsc}{1-\epsc} \:.
\ee

These results for spinless electrons can be directly carried over
to electrons with spin in the adiabatic regime,
when one corrects the Aharonov-Bohm phase $\theta_{\rm AB}$ by
the geometric phase
 \cite{diegoklaus}
\begin{equation}
\Gamma^{\suarr(\sdarr)} = \pi (1 + (-) \cos \alp),
\ee
following the replacement rule
\be \label{thetaabupdown}
\theta_{\rm AB} = \fr{\pi}{\phi_0} \phi^{\rm AB}
                ~\longrightarrow~
\theta_{\rm AB}^{\sigma} =   \fr{\pi}{\phi_0}
                \li( \phi^{\rm AB} - \phi_0 \fr{\Gamma^{\sigma}}{2 \pi} \re)
                \, ,
\ee
and accounts for the
Zeeman interaction energy in the dynamical phase,
\be \label{thetadupdown}
\theta_{d} = \pi \sqrteftil
               ~\longrightarrow~
\theta_{d}^{\suarr(\sdarr)} = \pi \sqrt{\eftil + (-)  \mutil B}
                \:.
\ee
The splitting of $\theta_{d}^{\suarr(\sdarr)}$ results in
interference and beating of the amplitudes of the
two electron species due to their slightly different oscillation frequencies,
which destroys the $\phi_0$-periodicity
of the Aharonov-Bohm effect (see also the discussion of
Fig.~\ref{fig_mm_fulltransm}).

\begin{figure}[!t]
\epsfxsize=7.5cm
  \centerline{\epsffile{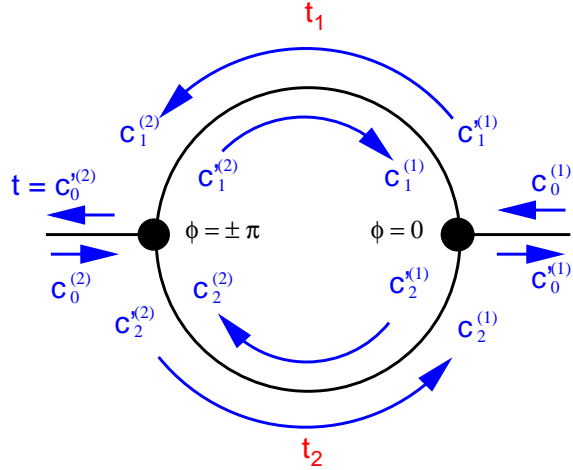}}
  \begin{center}
  \caption{Definition of the transmission and reflection amplitudes in a 1D
  ring coupled to current leads.
  \label{fig_yiamplitudes}
  }  \end{center}
\end{figure}


\subsection {Spin-dependent transport}
\label{subsec_spindep}

In spin-dependent transport, transitions between the two spin
channels $\spinup$, $\spindown$ are possible,
and the generalization of spin-independent transport requires
two changes.
Firstly, the wave functions acquire a spin dependence at each
spatial position, cf.~Sec.~\ref{sec_1dring},
and all transport equations have to be formulated in the
resulting product space (orbital motion $\otimes$ spin state).
This is explained in subsection~\ref{subsec_spindep_a}.
Secondly, at given total energy
 the electronic velocities $v_\rho^{\uarr,\darr}$
depend on the spin and propagation direction, because
the Zeeman energy is state-dependent, and also  the spin states (\ref{basisstates})
are not orthogonal, as already has been noted in Ref.~\onlinecite{yi}.
Here we find it necessary to depart from the derivation
of transfer coefficients in Ref.~\onlinecite{yi},
which resulted in non-unitary amplitude-relating
transfer matrices.
This would lead to a total dimensionless conductance $T$ that can
take values above two, in contradiction to particle number conservation.
If we wish to couple to the leads by the conventional unitary $S$ matrix (\ref{smatrix}),
we hence have
to work with suitable flux-normalized, orthogonal scattering states.
This is done in subsection~\ref{subsec_spindep_b}.

\subsubsection{Formalism}
\label{subsec_spindep_a}

The spin degree of freedom is incorporated formally by upgrading the coefficients ${\bf c}$ to
spinors.
Accordingly, they now consist of two
components
$\bf c^\uarr$, $\bf c^\darr$,
where $\uarr$ ($\darr$)
denotes the spin state in $z$-direction. The
transfer matrices $\matt_1,\matt_2$ are now $4 \times 4$ matrices, and
the transmission amplitude $t$ in Eq.~(\ref{tfromalp2}) is the $2 \times 2$
matrix
\be
\label{matt2mal2}
t = \li( \bma  t_{\uarr \uarr} & t_{\uarr \darr} \\
           t_{\darr \uarr} & t_{\darr \darr}
     \ema \re)
           \:.
\ee
The four entries of ${\matt}$
measure the transmission amplitudes between all
possible combinations of the spin states in the $S_z$-basis.

For unpolarized incident electrons, the
total dimensionless conductance $T$ is given by
\be \label{tprobsz}
T   =  \sum_{\sigma,\sigma'=\uarr,\darr} |t_{\sigma \sigma'}|^2
    =  T_{\uarr \uarr} + T_{\uarr \darr} +
                T_{\darr \uarr} +T_{\darr \darr} \:.
\ee

Equation (\ref{tfromalp2}) for the transmitted
amplitude is replaced by
\be\label{alp2yi}
    t    = - \epsc \, m_1 \, \matt_1 \, \matPi m_2  \:,
\ee
with
\bea
\matPi &=& \left[ \li( \matSre \otimes \sig_0 \re) \matt_{2}
         \li(\matSre  \otimes \sig_0 \re) \matt_1 -
         \openone_{2 \times 2} \otimes \sig_0 \right]^{-1}
         \:, \label{piyi} \\
m_1 &=& \fr{1}{b} \li( [1,1] \otimes \sig_0 \re) \nn \:, \quad
m_2 = \fr{1}{b} \li( \li[ {1 \atop -1} \re] \otimes \sig_0 \re)
\nn \:.
\eea
Here $\sigma_0$ is the $2\times 2$ unit matrix in spin space.
The flux-conserving transfer matrix $t_1$ in the upper arm
and its counterpart $t_{2}$ in the lower arm
are derived in the following Sec.~\ref{subsec_spindep_b}.

For perfectly symmetric rings we can derive a convenient simplified version of Eq.\ (\ref{alp2yi}).
To this end we introduce the matrix
\be\label{matm3}
\matmd = \fr{1}{b}
                \li(
                \begin{array}{cccc}
                a+b & 0 & a & 0 \\
                0 & a+b & 0 & a \\
                -a & 0 & 1 & 0 \\
                0 & -a & 0 & 1
                \end{array}
                \re)
                = \matNe \: \matNz \:,
\ee
with the decomposition
\be\label{matn1}
\matNe = \fr{1}{\sqrt{2 b}}
                \li(
                \begin{array}{cccc}
                b + a & 0 & -1 & 0 \\
                0 & b+a & 0 & -1 \\
                1 & 0 & 1 & 0 \\
                0 & 1 & 0 & 1
                \end{array} \re)\:,
\ee
\be\label{matn2}
\matNz = \fr{1}{\sqrt{2 b}}
                \li(
                \begin{array}{cccc}
                1 & 0 & 1 & 0 \\
                0 & 1 & 0 & 1 \\
                -(b+a) & 0 & 1 & 0 \\
                0 & -(b+a) & 0 & 1
                \end{array} \re) \:.
\ee
This allows us to rewrite
Eq.~(\ref{alp2yi}) in the symmetric form
\[
t = - \epsc \: m_1 \:  \matNz^{-1} \: \li[
                \matNz \matt_{2} \matNe  - \matNe^{-1} \matt_1^{-1}  \matNz^{-1}
                 \re]^{-1} \matNe^{-1} \: m_2 \nn \:.
\]
If we furthermore introduce
\be \label{mata}
\mata = \li[
                \matNz \matt_{2} \matNe  - \matNe^{-1} \matt_1^{-1}  \matNz^{-1}
                \re]^{-1} \:,
\ee
we see from the previous equation
that the term to the left and right of
$\mata$ projects out just the {\em upper right} $2 \times 2$ matrix of
$\mata$,
\be \label{matarechtsoben}
\matt =  - \epsc \:\fr{2}{b}
                        \li( \bma P'_{13} & P'_{14} \\
                                P'_{23}& P'_{24}
                         \ema \re) \:,
\ee
which simplifies the calculation
of the dimensionless conductance from
Eqs.~(\ref{matt2mal2}, \ref{tprobsz}).

\subsubsection{Computation of the transfer matrices}
\label{subsec_spindep_b}

We now turn to the computation of the flux-normalized transfer matrices $\matt_1, \matt_{2}$,
taking care for the
state dependence of the propagation velocities $v_\rho^{\uarr(\darr)}$
and for the non-orthogonality of the spin states (\ref{basisstates}).

In analogy to the
spinless case, we first introduce transfer matrices $\mattt_1, \mattt_2$ that relate
the amplitudes rather than the fluxes.
With the velocity matrix
\be
v = {\rm diag}(v_+^\uarr,v_+^\darr, v_-^\uarr, v_-^\darr)
\ee
the relation between $\mattt_1, \mattt_2$ and $\matt_1, \matt_{2}$
is given by
\be
\label{mf}
\matt_1  =  v^{1/2} \mattt_1 v^{-1/2} \:, \:
\matt_{2}  =  v^{1/2} \mattt_2 v^{-1/2}  \:.
\ee

Following the wave-matching procedure  for ballistic transport in
Ref.\ \onlinecite{yi}, the transfer matrix $\mattt_1$  in the basis of
non-flux-normalized  eigenfunctions (\ref{basisstates}) is
of the block-diagonal form
\be\label{matt1}
\mattt_{1} = \li(
  \begin{array}{cccc}
    g_1 &
    \tilde g_2 & 0 & 0 \\
    \tilde g_3 & g_4 & 0 & 0 \\
    0 & 0 & h_1 & \tilde h_2\\
    0 & 0 & \tilde h_3 & h_4
  \end{array} \re)\:.
\ee
With $\omega_\rho=\cos ( \zeta_\rho^{\suarr} - \zeta_\rho^{\sdarr} )$, where
$\zeta_\rho^\sigma = \gamma_\rho^\sigma/2$,
the entries for counter-clockwise propagation read
\bea
g_1 &=& \fr{1}{\omega_+}
        \li(
        e^{i n_+^{\suarr} \pi} \cos \zeta_+^{\suarr} \cos \zeta_+^{\sdarr}
       +e^{i n_+^{\sdarr} \pi} \sin \zeta_+^{\suarr} \sin \zeta_+^{\sdarr}
        \re)
        \:, \nn \\
\tilde g_2 &=& \fr{e^{- i \varphi_t} }{\omega_+}
        \li(
        e^{i n_+^{\suarr} \pi} - e^{i n_+^{\sdarr} \pi} \re)
        \cos \zeta_+^{\suarr} \sin \zeta_+^{\sdarr}
        \:, \nn \\
\tilde g_3 &=& \fr{- e^{i \varphi_t} }{\omega_+}
        \li(
        e^{i n_+^{\suarr} \pi} - e^{i n_+^{\sdarr} \pi} \re)
        \sin \zeta_+^{\suarr} \cos \zeta_+^{\sdarr}
        \label{g1g2g3g4} \:,\\
g_4 &=& \fr{- 1}{\omega_+}
        \li(
        e^{i n_+^{\suarr} \pi} \sin \zeta_+^{\suarr} \sin \zeta_+^{\sdarr}
       +e^{i n_+^{\sdarr} \pi} \cos \zeta_+^{\suarr} \cos \zeta_+^{\sdarr}
        \re) \nn \:.
\eea
Similar expressions apply to the clockwise propagating waves,
\bea
h_1 &=& \fr{1}{\omega_-}
        \li(
        e^{i n_-^{\suarr} \pi} \cos \zeta_-^{\suarr} \cos \zeta_-^{\sdarr}
       +e^{i n_-^{\sdarr} \pi} \sin \zeta_-^{\suarr} \sin \zeta_-^{\sdarr}
        \re)
        \:, \nn \\
\tilde h_2 &=& \fr{- e^{- i \varphi_t} }{\omega_-}
        \li(
        e^{i n_-^{\suarr} \pi} - e^{i n_-^{\sdarr} \pi} \re)
        \cos \zeta_-^{\suarr} \sin \zeta_-^{\sdarr}
        \:, \nn \\
\tilde h_3 &=& \fr{ e^{i \varphi_t} }{\omega_-}
        \li(
        e^{i n_-^{\suarr} \pi} - e^{i n_-^{\sdarr} \pi} \re)
        \sin \zeta_-^{\suarr} \cos \zeta_-^{\sdarr}
        \:, \label{h1h2h3h4}  \\
h_4 &=& \fr{- 1}{\omega_-}
        \li(
        e^{i n_-^{\suarr} \pi} \sin \zeta_-^{\suarr} \sin \zeta_-^{\sdarr}
       +e^{i n_-^{\sdarr} \pi} \cos \zeta_-^{\suarr} \cos \zeta_-^{\sdarr}
        \re)  \:. \nn
\eea

We obtain $\matt_1$ from relation (\ref{mf})
after introducing the square root of the velocity ratios
(obtained from the argument
of particle conservation, or through direct computation)
$\ny_\rho =  \sqrt{v_\rho^{\uarr} / v_\rho^{\darr}} =
\sqrt { \tan \zeta_\rho^{\suarr} / \tan \zeta_\rho^{\sdarr} }$
as
\be\label{mattI}
\matt_1 = \li(
  \begin{array}{cccc}
    g_1 &
    g_2 & 0 & 0 \\
    g_3 & g_4 & 0 & 0 \\
    0 & 0 & h_1 &  h_2\\
    0 & 0 & h_3 & h_4
  \end{array} \re)\:,
\ee
where $g_2=\ny_+ \tilde g_2$, $g_3 = \ny_+^{-1} \tilde g_3$,
$h_2=\ny_- \tilde h_2$, $h_3=\ny_-^{-1} \tilde h_3$.

The same algebra performed for the lower arm yields
\be\label{mattII}
\matt_2 = \li(
  \begin{array}{cccc}
    g_1 & - g_2 & 0 & 0 \\
    - g_3 & g_4 & 0 & 0 \\
    0 & 0 & h_1 &  - h_2\\
    0 & 0 & - h_3 & h_4
  \end{array} \re)\:.
\ee

Now we have provided all ingredients to calculate transmission
amplitudes and probabilities according to
Eqs.~(\ref{tprobsz},\ref{mata},\ref{matarechtsoben}) for
any desired field configuration that respects
rotational symmetry.

\subsubsection{Transformations in spin space}
So far, we always used the $S_z$-basis to express the spin states
needed to calculate the transfer matrices.
Here, the transmission probabilities $t_{\uarr \darr}, t_{\darr \uarr}$ for spin-flip processes
are non-zero for a tilted
magnetic field even in the adiabatic limit.
We recall that there are no transitions between the propagating
states $\sspinupphi$, $\sspindownphi$ in this case. This is, of
course, no contradiction since
$\sspinupphi$, $\sspindownphi$
represent spins aligned with the magnetic field, given in
$S_z$-basis by Eq.~(\ref{spineigalp_unif}).

Alternatively, one can consider transmission amplitudes in the
local $S_{\alp}$-basis (\ref{spineigalp_unif}).
The new transmission amplitudes
$t^{\alp}_{\suarr \suarr}, t^{\alp}_{\suarr \sdarr},
t^{\alp}_{\sdarr \suarr}, t^{\alp}_{\sdarr \sdarr}$
that replace the ones with respect to the $S_z$-basis in the
matrix $\matt$, Eq.~(\ref{matarechtsoben}),
are obtained by performing the appropriate projections, e.g.,
\be
\label{talpupup}
t^{\alp}_{\suarr \suarr} =
        \la \suarrbig(\varphi=\pi) |\matt |\suarrbig(\varphi=0)\ra
        \:.
\ee

In Fig.~\ref{fig_transmszsalp} we show an example for
the full as well as the partial transmission probabilities
in the $S_z$- and $S_\alpha$-basis.
Figure~\ref{fig_transmszsalp}(a) illustrates the oscillations in the
dimensionless conductance as function of the scaled Fermi momentum $\keffrr = \sqrteftil$
for parameters chosen in the adiabatic regime ($Q > 1$).
The diagonal and off-diagonal transmission probabilities
in $S_z$- and $S_{\alp}$-basis are shown in
Figs.~\ref{fig_transmszsalp}(b),(c) for electrons entering the
ring with initial spin parallel to the magnetic field
(state $|\suarr\rangle$).
Note the differences between the $S_z$- and $S_{\alp}$-representation ---
the spin-switching components, $T^{\alp}_{\suarr \sdarr},T^{\alp}_{\sdarr \suarr}$,
only vanish in $S_{\alp}$-basis as expected under adiabatic conditions.
We point out that the off-diagonal partial transmissions coincide
due to the reflection symmetry of the system about an axis perpendicular to the leads
through the center of the ring.


\section{Geometric phases in the magneto-conductance}
\label{sec_general}

Here and in the following section we discuss the influence of
geometric and Aharonov-Bohm phases on the magneto-conductance through the 1D rings.
In this Section we consider an inhomogeneous
magnetic field generated by a central micromagnet and study the
magneto-conductance as the Fermi energy of the incident electrons
or the magnetic field in
$z$-direction is varied by applying an external magnetic field.
In Sec.~\ref{sec_inplane}
we will concentrate on in-plane magnetic fields ($\alp \approx \pi/2$)
and use $B_z$ as a small control field.

\begin{figure}[!t]
  \epsfxsize=8.5cm
  \centerline{ \epsffile{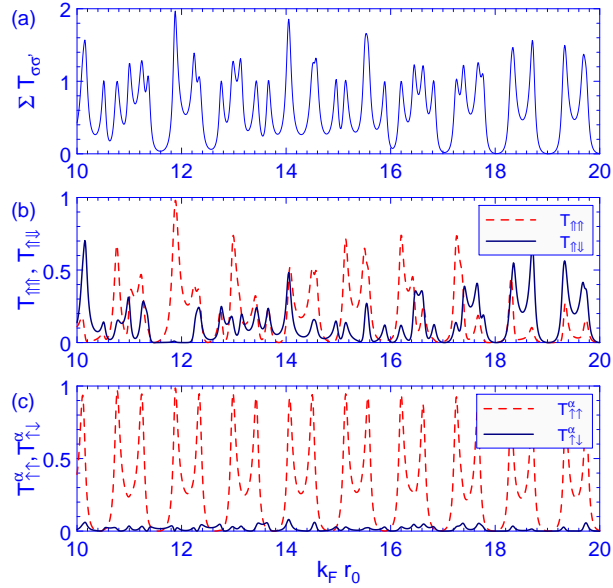}}
  \caption{(a) Dimensionless conductance $T=\sum_{\sigma,\sigma'} T_{\sigma\sigma'}$ and partial
  contributions in (b) the $S_z$-basis and (c) the $S_{\alp}$-basis
  vs.~the scaled Fermi momentum
  $\keffrr$ of the incident electrons
  ($g^*=1, m^*=m_0$, coupling strength $\epsc=0.25$, tilt angle $\alp = \arctan 2 \approx 63.4^o$,
  adiabaticity parameter between $Q=3.35$ (for $\keffrr=10$) and
  $Q=1.68$ (for $\keffrr=20$), filed texture
  parameter $\varphi_t= \pi/2$); corresponding to magnetic field components
  $B_r=0, \, \mutil B_{\varphi} = 30 \phi_0/(\pi r_0^2), \, \mutil B_z = 15 \phi_0/(\pi r_0^2)$.
  Shown are the diagonal and off-diagonal partial transmissions for electron
  initially in the state $|\suarr \rangle$,
  parallel to the field.
  For all values of $\keffrr$ we are in an
  (almost) adiabatic situation, as confirmed by the small spin-flip
  probabilities in $S_{\alp}$-basis in panel (c).
}
  \label{fig_transmszsalp}
\end{figure}

Our main motivation to study rotational invariant configurations of the
magnetic field is the experimental realization of a
crown-like magnetic field
by means of a central micromagnet as reported in Ref.~\onlinecite{ye}.
The magnetic field of the
cylinder-shaped Dysprosium micromagnet can be approximated by a dipole field at
the position of the ring. \cite{dissmh} It contains both a radial and a
$z$-component, $\vec{B}_{\rm M} = B_r \er + B_{z, \rm M} \ez$. The total magnetic field is
obtained after adding the homogeneous magnetic field $\vec{B}_{\rm ext} = B_{\rm ext} \ez$ in
$z$-direction, $\vec{B} = \vec{B}_{\rm M} + \vec{B}_{\rm ext}$. Both $\vec{B}_{\rm M}$ and
$\vec{B}_{\rm ext}$ contribute to the Aharonov-Bohm flux through the ring, and we denote
the magnetic flux due to the homogeneous field by $\phi^{\rm AB}_{\rm ext} = \pi r_0^2
B_{\rm ext}$.
The strength of the magnetic field depends on the premagnetisation
procedure, but is constant during the magneto-conductance measurement where the
homogeneous magnetic field perpendicular to the ring is varied. The
magnetization of the micromagnet is
characterized by the parameters
\be
Q_M = \fr{\mutil |\vec{B}_M|}{\keffrr} \:,\quad \quad
M^{\rm AB} = - \fr{\pi \rr^2 B_{z,\rm M}}{\phi_0} \:, \label{defmab}
\ee
where $Q_M$ is introduced in analogy to Eq.~(\ref{defQ}) as
adiabaticity parameter for the micromagnet for vanishing external flux,
$\phi^{\rm AB}_{\rm ext} = 0$.

The results for the calculated magneto-conductance are shown in
Fig.~\ref{fig_mm_fulltransm} for three different degrees of adiabaticity, $Q_M
=$ 0.4 [Fig.~\ref{fig_mm_fulltransm}(b)], 1 [Fig.~\ref{fig_mm_fulltransm}(c)],
10 [Fig.~\ref{fig_mm_fulltransm}(d)]. This parameter is adjusted by the proper
choice of the effective gyromagnetic ratio $g^*$ and the effective mass $\meff$,
leaving the strength of the micromagnet constant at $M^{\rm AB}$ = 5. The
connection to a specific material system is then provided by choosing the value
for $Q_M$ that corresponds to the product $g^* \meff$. We follow the geometry
described in Ref.\ \onlinecite{ye} and choose the radius of the ring as $\rr =
500 \, nm$. The 2DEG is placed in a plane lying $150 \, nm$ above the central
plane of the Dysprosium. We assume maximal coupling between ring and leads,
$\epsc$ = 0.5.

For comparison, we show in Fig.~\ref{fig_mm_fulltransm}(a)
the result of B\"uttiker {\it et al.}~\cite{imry}
for spin-independent transport,
with the
well-known Aharonov-Bohm oscillations
as the homogeneous magnetic field is varied.
However, here
we have taken into account the Zeeman splitting of the
energy and the influence of the Berry phase (assuming an adiabatic
situation), as is discussed at the end of Sec.\ \ref{subsec_spinindep}.
In Fig.~\ref{fig_mm_fulltransm}(b)-(d)
the strongest deviations from Aharonov-Bohm-like oscillations
are seen around $\phi^{\rm AB}_{\rm ext} / \phi_0 = M^{\rm AB} = 5$, indicating
the importance of geometric phases there. In fact, this corresponds to the
situation where the external flux $\phi^{\rm AB}_{\rm ext}$ cancels the
magnetic flux due
to the micromagnet such that the nonuniform field of the
micromagnet becomes maximally important.
The strong interference effects around
$\phi^{\rm AB}_{\rm ext} / \phi_0 = M^{\rm AB}$
stem from the slightly different
oscillation frequencies of the $\sspinup$- and
$\sspindown$-electrons due to the Berry phase.
Regular Aharonov-Bohm-like
oscillations are recovered as the external magnetic field
$B_{\rm ext}$ becomes dominant, which requires a smaller value of this field
in the diabatic situation
of Fig.~\ref{fig_mm_fulltransm}(b), when compared to the intermediate case of
Fig.~\ref{fig_mm_fulltransm}(c)
or to the  adiabatic case of Fig.~\ref{fig_mm_fulltransm}(d).
Note that the curves in Fig.~\ref{fig_mm_fulltransm}(b)-(d) are not quite
symmetric
about the line $\phi^{\rm AB}_{\rm ext}/\phi_0 = M^{\rm AB}$ (corresponding to a vanishing
overall Aharonov-Bohm flux $\phi^{\rm AB}=0$), since there
the angles $\gamma_\rho^\sigma$ still change monotonically.

We point out that there exist two
adiabatic limits: one for dominating field of the
micromagnet (parameterized by $Q_M$),
the other one for dominating external field $B_{\rm ext}$, which is always reached
for sufficiently large $|\phi^{\rm AB}_{\rm ext}| \gg |M^{\rm AB}| \phi_0$.
In  the sequence of Figs.~\ref{fig_mm_fulltransm}(b)-(d),
the adiabaticity  is increased with respect to the first limit, i.e. the
magnetic energy due to the micromagnet (at zero external flux $\phi^{\rm AB}_{\rm ext}$)
becomes large in comparison to the Fermi energy of the electrons.
As a result, the magneto-conductance in Fig.~\ref{fig_mm_fulltransm}(d) is well
described by the sum of the two curves
for the electron gases $||\suarr\ra\ra$ and $||\sdarr\ra\ra$
in Fig.~\ref{fig_mm_fulltransm}(a), which is indicated by the
dashed curve in Fig.~\ref{fig_mm_fulltransm}(d).

The above-mentioned experiment \cite{ye} was performed under rather diabatic
conditions [similar to those of Fig.~\ref{fig_mm_fulltransm}(b)], which are not
favorable for the detection of geometric phase effects. Accordingly, it was
found that the experimental observations could not be accounted for by geometric
phases. However, evidence for geometric phase effects in electronic ring
structures should be possible in more adiabatic regimes, that could, e.g., be
achieved
using stronger micromagnets,
appropriately arranged ferromagnetic particles, \cite{NBH00} or by exploiting
the spin-orbit interaction. \cite{yang,yau}
The magneto-conductance measured in the experiment of Ref.~\onlinecite{yang}
for a singly-connected InAs ring with a radius 250 $nm$ and a spin-orbit
induced magnetic field strongly resembles our results presented in
Fig..~\ref{fig_mm_fulltransm}(c),(d).

\begin{figure}[!t]
  \epsfxsize=8.5cm
  \centerline{\epsffile{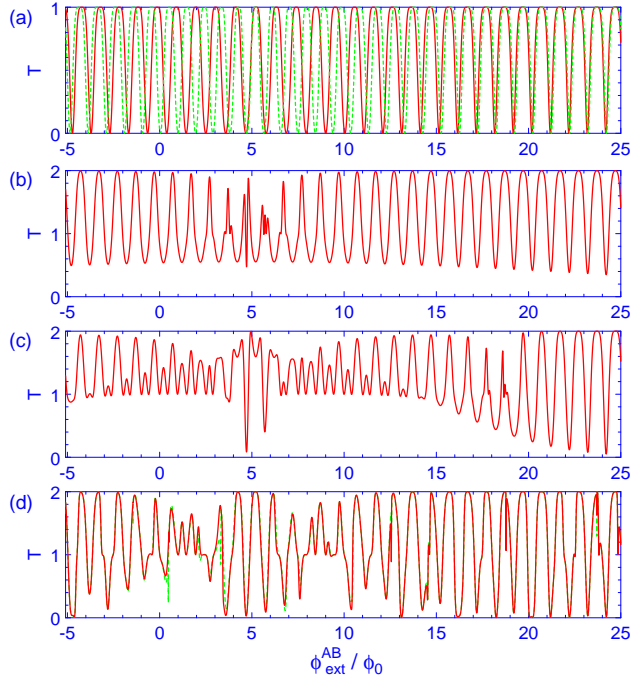}}
  \caption{  Magneto-conductance for an Aharonov-Bohm device with a
  central micromagnet ($\varphi_t=0$, $M^{\rm AB} = 5$)
  under variation of the external flux
  $\phi^{\rm AB}_{\rm ext}$. In (a) we show the dimensionless conductance
  for independent electron gases $||\suarr\ra\ra$ (solid curve) and
  $||\sdarr\ra\ra$ (dashed curve)
  as in Ref.~\protect\onlinecite{imry}, but taking the Zeeman
  energy and Berry phase into account.
  In (d), the sum of the two contributions
  is shown as dashed line. The dimensionless conductance for
  spinful electrons is given in panels (b)-(d);
  (b) diabatic regime ($Q_M=0.4$),
  (c) intermediate
  case ($Q_M=1$) where effects due to geometrical phases become visible,
  (d) adiabatic limit ($Q_M=10$), which is dominated by interference effects
  due to different Berry phases for the
  electron gases $||\suarr\ra\ra$ and $||\sdarr\ra\ra$. We point out the
  similarity to the magneto-conductance {\it measured} in the experiment
  of Ref.~\protect\onlinecite{yang}.
  The effect
  of geometric phases is always lost for dominating external field,
  $\phi^{\rm AB}_{\rm ext} \gg M^{\rm AB} \phi_0$ because
  then $\alp \rightarrow 0$.
}
  \label{fig_mm_fulltransm}
\end{figure}


\section{In-plane magnetic field}
\label{sec_inplane}

\subsection{Aharonov-Bohm ring as a spin-switch}
\label{subsec_spinswitch}

We will now consider a ring that is subject to an in-plane magnetic field, which
can be either circular, radial, or a rotationally invariant combination of the
two. In experiments, such a field could be generated, e.g., by a current
through the ring.
\cite{current}
We recently reported
a {\em spin-flip effect}  for this magnetic-field texture,
\cite{2dlang,prl} by which
one can change the spin polarization
of the transmitted electrons
with a small (external) Aharonov-Bohm flux
through the ring. The spin-dependent transmittance is periodic in the applied
magnetic flux, with a period of one flux quantum $\phi_0$.
In particular, the polarization
state of polarized electrons can be inverted by altering
this magnetic control flux
by $\phi_0 /2$, which might become interesting
for future spintronic devices. In paper II of this series (Ref.\ \onlinecite{2dlang})
we investigated in detail
necessary conditions for the spin-flip effect to hold.

For an in-plane magnetic field, $\alp = \pi / 2$,
we find a high symmetry between the clockwise and
counter-clockwise propagating wave. This becomes manifest in
simple relations between the angles $\gamma_\rho^\sigma$ entering the
spin states (\ref{basisstates}) and the orbital
quantum numbers $n_\rho^\sigma$ which solve Eq.~(\ref{polynomnpr}). For the case
with no
magnetic flux through the ring, $\phi^{\rm AB} = 0$, we find
\bea
 \gamma_+^{\suarr} = \gamma_-^{\sdarr} \:,
 & \quad &
 n_+^{\suarr} = -n_-^{\sdarr} - 1 \:, \nn \\
 \gamma_+^{\sdarr} = \gamma_-^{\suarr} \:,
 & \quad &
 n_+^{\sdarr} = -n_-^{\suarr} - 1 \:.
\eea
This simplifies
the structure of the transfer matrices
$\matt_1, \matt_{2}$, cf.~Eqs.~(\ref{mattI}, \ref{mattII}),
since we can now use the relations
\bea
g_1 = h_4^* & \:,\quad & g_2 = h_3^* \:, \nn \\
g_4 = h_1^* & \:,\quad & g_3 = h_2^* \:. \label{gsundhsphi0}
\eea

The number of variables can be reduced further if we specify
the field texture parameter $\varphi_t$
in order to relate $g_2$ and $g_3$.
We find $g_2 =- (+) g_3$
in the case of a radial (circular) magnetic field, $\varphi_t = 0$ ($\varphi_t =
\pi/2$).
These relations still hold when an integer number of flux quanta
penetrate the ring,
as can be seen from Eqs.~(\ref{g1g2g3g4}, \ref{h1h2h3h4}),
because this corresponds to changes of $n$ by an integer---and an
overall sign change, where applicable, does not change the
transmission amplitude.
Together with the definitions
(\ref{matm3}-\ref{matn2}, \ref{mattI}, \ref{mattII}), the
relations in (\ref{gsundhsphi0}) and the unitarity and symmetry
of the matrices $\matt_1, \matt_{2}$,
we now can give a rigorous analytical proof of the
above-mentioned spin-switch effect.

Evaluating the matrix $\mata^{-1}$ given by Eq.~(\ref{mata})
in a straightforward computation for $\phi^{\rm AB} / \phi_0
\in \mathbb {N}$, we obtain for a circular magnetic field, $g_2 =
g_3$,
\[
\mata^{-1} = \fr{1}{2 b} \li(
  \begin{array}{cccc}
    \delta_1 & \delta_2 & \delta_3 & 0 \\
    \delta_2 & -\delta_1 & 0 & \delta_3 \\
    \delta_4 & 0 & \delta_1 & \delta_2 \\
    0 & \delta_4 & \delta_2 & -\delta_1
  \end{array}
  \re) \:,
\]
with
\bea
\delta_1 & = & (g_1 - g_4) (a+b) - (g_1^* - g_4^*) \:, \nn \\
\delta_2 & = & - 2 \li( g_2 (a+b) +  g_3^* \re) \:, \nn \\
\delta_3 & = & -(g_1 + g_4) + (g_1^* + g_4^*) \:, \nn \\
\delta_4 & = & -(g_1 + g_4) (a+b)^2 + (g_1^* + g_4^*) \label{deltas} \:.
\eea
The inversion of this matrix yields
\bea
P'_{13} = P'_{24} & = &
 2 b \fr{\delta_3}{\delta_3 \delta_4 - \li(\delta_1^2 + \delta_2^2\re)}
 \:, \nn \\
P'_{14} = P'_{23} & = & 0 \:, \label{a14a23noflux}
\eea
and these results carry over
to the transmission amplitudes (\ref{matarechtsoben}) in the $S_z$-basis,
\[
t_{\uarr \uarr} = t_{\darr \darr} \quad \textrm{and} \quad
t_{\uarr \darr} = t_{\darr \uarr} = 0 \:.
\]
We also  can express the
transmission probabilities
in the $S_{\alp}$-basis according to
Eq.~(\ref{talpupup}),
yielding
\be
t^{\alp}_{\suarr \suarr} = t^{\alp}_{\sdarr \sdarr}
= t_{\uarr \darr} = t_{\darr \uarr} = 0  \:, \;
t^{\alp}_{\suarr \sdarr} = t^{\alp}_{\sdarr \suarr}
= t_{\uarr \uarr} = t_{\darr \darr} \:,
\label{talpplanenull}
\ee
that is, the role of the diagonal ($\uarr \uarr$ and $\darr \darr$)
and non-diagonal elements ($\uarr \darr$ and $\darr \uarr$)
is just interchanged when switching between $S_z$- and $S_\alp$-basis.

The result (\ref{talpplanenull})
states that only electrons that {\em flip} \cite{remark_spinrichtung}
their spin
with respect to the magnetic field
are transmitted through the ring.
Moreover, the properties (\ref{talpplanenull}) of the transmission
amplitudes remain valid for any in-plane magnetic field configuration.
This explains the numerical findings for 2D rings in Refs.\ \onlinecite{2dlang,prl}.
Note that the spin-flip effect holds also
in the adiabatic limit. Since this counteracts the adiabatic spin alignment,
the dimensionless conductance tends to zero in this limit.

We now turn to the case where half a flux quantum penetrates the ring,
such that there is an Aharonov-Bohm flux $\phi^{\rm AB} = \phi_0 / 2$.
We assume the magnetic field at the position of the ring to remain
unchanged, as in the case of an
AB flux generated by a solenoid (rather than by a homogeneous
$B_z$-component).
The quantum  number $n$
is then reduced by half an integer, cf.~Eq.~(\ref{nprn}),
and Eq.~(\ref{gsundhsphi0}) has to be
replaced by
\bea
g_1 = - h_4^* & \:,\quad & g_2 = - h_3^* \:, \nn \\
g_4 = - h_1^* & \:,\quad & g_3 = - h_2^* \:. \label{gsundhsphihalbe}
\eea
These relations also directly follow from
Eqs.~(\ref{g1g2g3g4}, \ref{h1h2h3h4}), since the coefficients are multiplied by a factor
$e^{-i \pi/2}$ compared to the values obtained for $\phi^{\rm AB} = 0$.
Hence we can write
\be\label{matt1withhalfflux}
\matt_{j}^{\phi_0/2} = - i \matt_{j}^0\:,
\ee
where $\matt_{j}^0$ are the transfer matrices for $\phi^{\rm AB} =0$.
Accordingly, we now find
\bea
{\mata}(\phi_0/2) & = & \li[
                -i \matNz {\matt}_{2}^0 \matNe  - \fr{1}{-i}
                \matNe^{-1} {\matt_1^0}^{-1}  \matNz^{-1}
                 \re]^{-1} \:, \nn \\
& = & - \fr{1}{i}\li[
                 \matNz {\matt}_{2}^0 \matNe  +
                \matNe^{-1} {\matt_1^0}^{-1}  \matNz^{-1} \re]^{-1}
                \label{mataphi0halbe} \:.
\eea
For a circular magnetic field ($g_2 = g_3$) this results in
\[
\li({\mata}(\phi_0/2)\re)^{-1} = \fr{1}{2 b} \li(
  \begin{array}{cccc}
    \delta'_1 &  0  & \delta'_2& \delta'_3\\
    0 & -\delta'_1 & \delta'_3& -\delta'_2\\
    \delta'_4& \delta'_5 & \delta'_1 & 0 \\
    \delta'_5 & -\delta'_4& 0 & -\delta'_1
  \end{array}
  \re) \:,
\]
with
\bea
\delta'_1 & = & (g_1 + g_4) (a+b) + (g_1^* + g_4^*) \:, \nn \\
\delta'_2 & = & -(g_1 - g_4) - (g_1^* - g_4^*) \:, \nn \\
\delta'_3 & = &  2 \li( g_2 - g_3^* \re) \:, \nn \\
\delta'_4 & = & -(g_1 - g_4) (a+b)^2 - (g_1^* - g_4^*) \:, \nn \\
\delta'_5 & = & 2 \li( g_2 (a+b)^2 - g_3^* \re) \label{rhos} \:,
\eea
where the $g_i$
are the same as for  $\phi^{\rm AB} = 0 $.
Inverting this matrix one obtains
\bea
P_{13}^{\prime}(\phi_0/2) = - P_{24}^{\prime}(\phi_0/2) & = &
 2 b \fr{- \delta_1^{\prime 2} \delta'_2 +  \li(\delta_2^{\prime 2} + \delta_3^{\prime 2}\re) \delta'_4} {\cal N} \nn \\
P_{14}^{\prime}(\phi_0/2) = P_{23}^{\prime}(\phi_0/2) & = &
 2 b \fr{- \delta_1^{\prime 2} \delta'_3 +  \li(\delta_2^{\prime 2} + \delta_3^{\prime 2}\re) \delta'_5} {\cal N}
 \:, \nn
\eea
with the common denominator
\[
{\cal N} =  \delta_1^{\prime 4}
                + \li(\delta_2^{\prime 2} + \delta_3^{\prime 2}\re) \li(\delta_4^{\prime 2} + \delta_5^{\prime 2}\re)
                - 2 \delta_1^{\prime 2} (\delta'_2 \delta'_4 +  \delta'_3 \delta'_5)
                \:.
\]
Again, this carries over to the transmission amplitudes (\ref{matarechtsoben}), and the
result in $S_z$-basis reads
\[
t_{\uarr \uarr}^{\phi_0/2} = - t_{\darr \darr}^{\phi_0/2}
 \quad \textrm{and} \quad
t_{\uarr \darr}^{\phi_0/2} = t_{\darr \uarr}^{\phi_0/2}  \:,
\]
where all quantities are non-zero, in contrast to the result for the
non-diagonal transmission amplitudes in the case  $\phi^{\rm AB} =0$.
Transformation to the $S_{\alp}$-basis, however, yields
\be
{{}^{\alp}t^{\phi_0/2}_{\suarr \suarr, \sdarr \sdarr}} \neq 0, \quad
{{}^{\alp}t^{\phi_0/2}_{\suarr \sdarr, \sdarr \suarr}} = 0  \;\:.
 \label{talpplanephihal}
\ee
This result applies whenever the Aharonov-Bohm flux equals half an integer
number of flux quanta $\phi_0$, i.e. $\phi^{\rm AB} =\pm\phi_0/2, \pm 3
\phi_0/2, \ldots $.
The physical meaning of this result is that in the presence of half a
flux quantum only electrons that {\em keep} their spin direction
with respect to the magnetic field
during
transport are transmitted -- precisely the opposite of what we
found for $\phi^{\rm AB}=0$.
This opens the possibility of
controlling the transmission of (polarized) electrons by varying
the number of flux quanta penetrating the ring:
For  $\phi^{\rm AB} / \phi_0 $ integer,
only spin-flipping electrons are transmitted, whereas for
half-integer number of flux quanta only
electrons keeping their spin polarization can be found in the exiting lead.
For an illustration of these two situations see Figs.~\ref{fig_spinflip}(a) and (c).
For magnetic fluxes in between, the situation is intermediate with
transport in all channels,
see Fig.~\ref{fig_spinflip}(b) with $\phi^{\rm AB} = \phi_0 /4  $.

Remember that in the derivation of the spin-flip effect we have assumed the Aharonov-Bohm flux to
cause no change in the angle $\alp$ of the magnetic field $\vec{B}$ with the
$z$-axis. Therefore, in realistic situations where the magnetic control flux might originate from
a homogeneous magnetic field component $B_z$ in $z$-direction,
which also acts on the electronic spin in the ring,
we have to expect deviations from
the result (\ref{talpplanephihal}).
In order to test the sensitivity of the spin-flip effect,
we indeed followed this procedure in Fig.~\ref{fig_spinflip},
and find that for large in-plane magnetic fields used there the
influence of $B_z$ is negligible.

\begin{figure}[!t]
  \epsfxsize=8.5cm
  \centerline{ \epsffile{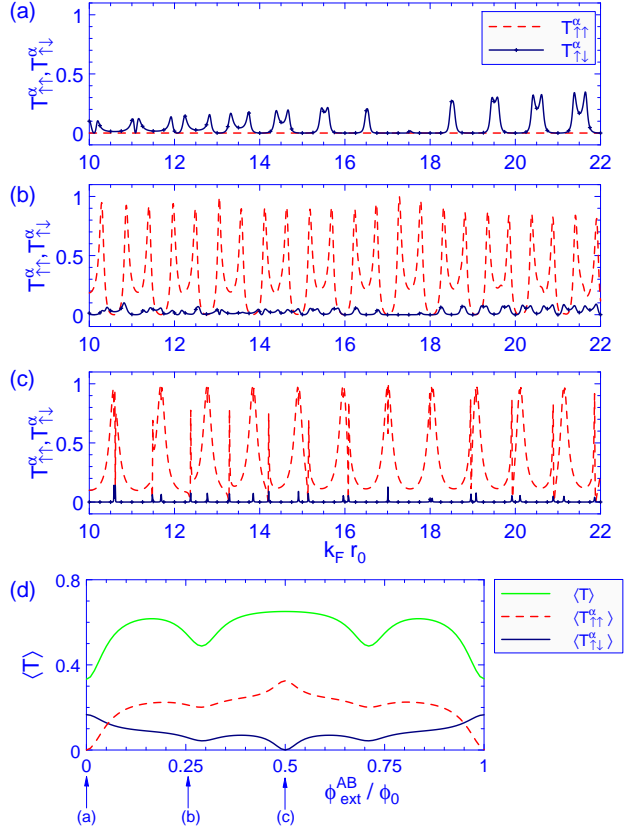}}
  \begin{center}
  \caption{
  Partial transmission probabilities
  vs.\ scaled momentum $\keffrr$ of the incident electrons
  ($g^* = 1, \meff = m_0, \epsc =0.25$)
  at fixed strength of the circular magnetic field
  $\mutil B_{\varphi} = 30 \phi_0/(\pi r_0^2)$ ($Q \approx 2$).
  (a)  $\phi^{\rm AB}$=0:
  (nearly) zero transmission for the smallest
  $\keffrr$ shown (due to the Berry phase in the
  adiabatic situation)
  and around $\keffrr = \mutil B / \sqrt{3} \approx 17.32 $.
  Note that the transmission is provided {\em only} by the
  spin-flipping channels, i.e.,
  only $T^{\alp}_{\suarr \sdarr}$ and
  $T^{\alp}_{\sdarr \suarr}$ are non-zero.
  (c)   $\phi^{\rm AB} =\phi_0/2$:
  opposite situation
  such that now $T^{\alp}_{\suarr \suarr} $ and $T^{\alp}_{\sdarr \sdarr}$ are
  non-zero.
  Finite values for the spin-flipping channels arise because
  the magnetic flux was created by a homogeneous field $B_z$
  affecting the angle $\alp$. (b) For intermediate values of the
  Aharonov-Bohm flux (here, $\phi^{\rm AB}=\phi_0/4$),
  transport occurs in all channels. Hence, for
  polarized incident electrons, the spin can be switched by applying
  an Aharonov-Bohm flux of half a flux quantum. This is shown in (d) for the
  energy-averaged transmission vs. control flux. The situations shown in (a)-(c)
  are marked by arrows.
  \label{fig_spinflip}
  }  \end{center}
\end{figure}

\subsection{Energy-averaged conductance
\label{subsec_avgt}}

So far we discussed the spin-flip effect
in terms of rapidly oscillating (total and partial) dimensionless
conductances.
Energy-averaging these quantities, while
keeping the degree of adiabaticity $Q$ constant,
cancels out the oscillations and reveals inherent features
that we will study in the following.

For an example of the averaged  dimensionless
conductance
at fixed $Q$ and variable Aharonov-Bohm flux $\phi^{\rm AB}$ see Fig.~\ref{fig_spinflip}(d).
The systematic dependence
on this parameter and the coupling strength $\varepsilon$ for $\phi^{\rm AB}=0$
is shown in Fig.~\ref{fig_3dinsetnoflux}. In the adiabatic limit ($Q \to \infty$),
the dimensionless conductance
approaches zero due to the effect of the Berry phase.
In the other, diabatic, limit ($Q \to 0$), the numerical data
agree with the 
analytical result (\ref{avgtvonepsana}) that was
multiplied by two to account for two open channels.

Most interestingly,
there are regions where $\la T \ra$
drops to zero that are not caused by geometric-phase effects.
In order to investigate what happens there, we look at the
non-vanishing transmission amplitudes (\ref{a14a23noflux}), in
particular at the numerator $\delta_3 = g_1^* - g_1 + g_4^* -
g_4 = -2\, i \, {\rm Im} (g_1 +g_4)$.
With the expressions for $g_1, g_4$ from Eq.~(\ref{g1g2g3g4}),
we find
\be
\label{delta3zeros}
\delta_3 = -2 \, i \: (\sin n_+^{\suarr} \pi - \sin n_+^{\sdarr} \pi)
           \: \fr{\cos (\zeta_+^{\suarr} + \zeta_+^{\sdarr})}
                {\cos (\zeta_+^{\suarr} - \zeta_+^{\sdarr})}\:,
\ee
implying that $\delta_3$ goes to zero whenever
the difference in the sine
factor vanishes. Obviously, this occurs when
$n_+^{\suarr}$ and $n_+^{\sdarr}$ differ by an even
integer number.
Expanding the difference $\Delta n=n_+^{\suarr}-n_+^{\sdarr}$
for small $\mutil B / \eftil $,
and using $\gam_+^{\suarr} \approx \gam_+^{\sdarr}$, we have to
solve the (Diophantic-like) equation
\[
\Delta n = \cos \fr{\gam_+^{\suarr} + \gam_+^{\sdarr}}{2}
Q
           \sin \fr{\gam_+^{\suarr} + \gam_+^{\sdarr}}{2}
\]
to yield $\Delta n$ as an even integer --- keeping in mind
that the $\gamma_\rho^\sigma$ themselves depend on the $n$.
The first zero is related to  $\Delta n = 2$
with $(\gam_+^{\suarr} + \gam_+^{\sdarr})/2 = \pi/3$,
giving
$Q = \sqrt{3}$
in accordance with the observation in Fig.\ \ref{fig_3dinsetnoflux}. This value is
consistent with the condition
$\mutil B / \eftil\ll 1$ as long as $\sqrteftil \gg \sqrt{3}$.
The further zeros in the averaged dimensionless conductance  belong
to $\Delta n = 2m$ (with the integer $m\ge 2$) and occur at ratios
$Q = \sqrt{4 m^2 -1}$, as is clearly visible in
Fig.~\ref{fig_3dinsetnoflux}.
Also note that the last factor in
Eq.\ (\ref{delta3zeros}) accounts for the overall decay
of $\la T \ra (Q)$, whereas the factor discussed
above is responsible for the oscillations with respect to $Q$.
The $Q$-dependence of
$\la T \ra$ around the zeros also strongly
depends on the coupling parameter $\epsc$,
see Figs.~\ref{fig_3dinsetnoflux} and \ref{fig_avgT_p5}.

\begin{figure}[!t]
  \epsfxsize=8.5cm
  \centerline{\epsffile{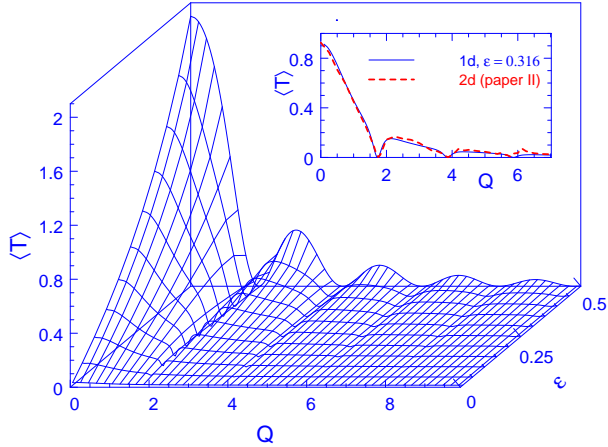}}
  \begin{center}
  \caption{
  Energy-averaged dimensionless conductance
  as a function of the adiabaticity parameter $Q$ for an in-plane
  circular  magnetic field ($\phi_t=\alpha=\pi/2$ ) and vanishing
  magnetic flux $\phi^{\rm AB} = \phi^{\rm AB}_{\rm ext} =0$.
  The tilted axis is the
  coupling parameter $\epsc$.
  The material-specific parameters $g^*$ for the gyromagnetic
  ratio and the effective mass $\meff$ are chosen here and in the
  following figures as $g^* = 15$ and
  $\meff = 0.023 \, m_0$ corresponding to InAs.
  The maximum $\la T \ra_{\rm max}$
  of $\la T \ra$ is observed at $Q = 0$,
  i.e., in the diabatic limit. It only depends on $\epsc$ and obeys the
  law $\la T \ra_{\rm max} = 2 \,\epsc / (1-\epsc) $ where
  $0 \le \epsc \le 0.5$. Minima in $\la T \ra$ occur at
  $Q= \sqrt{4 m^2 -1}$ ($m \in \mathbb {N}$). The overall decay of
  $\la T \ra$ to zero in the  adiabatic limit
  $Q \rightarrow \infty$ is an effect of the Berry
  phase. The inset shows a comparison between the results of the 2D model
  used in paper II  and the 1D model with $\epsilon=0.316$.
  \label{fig_3dinsetnoflux} }
  \end{center}
\end{figure}

We will now derive
an approximate analytical expression for the averaged
dimensionless conductance $\la T \ra$
for the case of maximal coupling strength $\epsc = 0.5$.
Equation (\ref{deltas}) yields
\bea
\delta_1 & = & - (g_1^* - g_4^*) \:, \nn \\
\delta_2 & = & - 2 g_3^* \:, \nn \\
\delta_3 & = & g_1^* - g_1 + g_4^* - g_4  \:, \nn \\
\delta_4 & = & g_1^* + g_4^* \:. \label{deltasepseinhalb}
\eea
For the denominator
${\cal N} \equiv \delta_3 \delta_4 - \li(\delta_1^2 + \delta_2^2\re)$
in Eq.~(\ref{a14a23noflux}), we obtain\be
{\cal N}
          =  - (g_1 + g_4) (g_1^* + g_4^*) + 4 {  \det g^*} \:,
\ee
with $\det g = g_1 g_4 -  g_3^2$.
Expressing the $g_i$ through Eqs.~(\ref{g1g2g3g4}),
and substituting
$\Delta n = n_+^{\suarr} - n_+^{\sdarr}$, we arrive at
\bea
(g_1 + g_4) (g_1^* + g_4^*)
      &=&  2 \fr{\cos^2 (\zeta_+^{\suarr} + \zeta_+^{\sdarr})}
                {\cos^2 (\zeta_+^{\suarr} - \zeta_+^{\sdarr})}
        \li( 1 - \cos \Delta n \pi \re) \:, \nn \\
\det g & = &    - e^{i ( n_+^{\suarr} + n_+^{\sdarr} ) \pi} \:. \nn
\eea
This finally leads to the analytical expression
for the dimensionless conductance (\ref{tprobsz})
for $\epsc = 0.5$:
\bea
T_{0.5}^{\rm ana} &=& T_{\uarr \uarr} + T_{\darr \darr}
 = \fr{2} {b^2} \: P_{13}^\prime P_{13}^{\prime *} \nn \\
 &=& \fr {8 \, \delta_3 \delta_3^*}
 {\li( |g_1 + g_4| ^2 + 4 \,{\det} g^* \re)
        \li( |g_1 + g_4| ^2 +  4 \,{\det g} \re) } \label{tanazwerg} \:.
\eea
In Eq.~(\ref{delta3zeros}) we already evaluated part of the numerator
that finally yields
\be
\delta_3 \delta_3^*
 =  16 \: \fr{\cos^2 (\zeta_+^{\suarr} + \zeta_+^{\sdarr})}
                {\cos^2 (\zeta_+^{\suarr} - \zeta_+^{\sdarr})}
              \:   \sin^2 \fr {\Delta n \pi}{2}
              \:   \cos^2 \fr{(n_+^{\suarr} + n_+^{\sdarr}) \pi}{2} \:.
                 \label{zaehT}
\ee
Furthermore, the denominator
of Eq.~(\ref{tanazwerg}) evaluates to
\bea
&  & 4 \, \fr{\cos^4 (\zeta_+^{\suarr} + \zeta_+^{\sdarr})}
       {\cos^4 (\zeta_+^{\suarr} - \zeta_+^{\sdarr})}
       \: (1-\cos \Delta n \pi)^2 + 16 \label{nennT}  \\
& - &      16 \, \fr{\cos^2 (\zeta_+^{\suarr} + \zeta_+^{\sdarr})}
       {\cos^2 (\zeta_+^{\suarr} - \zeta_+^{\sdarr})}
       \: (1-\cos \Delta n \pi)
       \cos (n_+^{\suarr} + n_+^{\sdarr}) \pi \:, \nn
\eea
where we have used the unitarity of $\matt_1,
\matt_{2}$ (${\det g \, \det g^*} = 1$).

The next step is to average Eqs.~(\ref{zaehT}, \ref{nennT}) over energy.
The result for the averaged dimensionless conductance is conveniently expressed after
introducing
the mean angle
$\bar{\gamma} = \fr{1}{2} (\gamma_+^{\suarr}+\gamma_+^{\sdarr})$,
the mean kinetic energy quantum number $\bar{n} = \fr{1}{2} ( n_+^{\suarr} +
n_+^{\sdarr} )$, and the difference angle $\Delta \gamma =
\gamma_+^{\suarr} - \gamma_+^{\sdarr}$. Then Eq.~(\ref{gamma1}) reads
\[
\cot \bar{\gamma} = \fr{\bar{n} + \fr{1}{2} } {\mutil B} \:,
\]
and we obtain the relations
\bea
\bar{n} &=& \keffrr - \fr{1}{2} +
              \sin \fr{\Delta \gamma}{2} \sin \bar{\gamma}
              \approx \keffrr - \fr{1}{2} \:, \nn \\
\Delta n &=& \cos \bar{\gamma} +
                \fr{\mutil B}{\keffrr} \sin \bar{\gamma}
                \:,\nn \\
\tan \Delta \gamma &=& - \fr{\Delta n}
                        {\mutil B + \fr{\bar{n} (\bar{n}-1)}{\mutil B}}
                \:. \nn
\eea
Furthermore, we can express $\bar{\gamma}$ and $\Delta n $
in terms of the adiabaticity parameter
$Q$ defined in Eq.~(\ref{defQ}),
which elucidates the role of the geometric phase as a measure of
adiabaticity,
\be
\cos \bar{\gamma} = \fr{1}{\sqrt{1+Q^2}}\:, \quad \quad
\Delta n = \sqrt{1+Q^2}  \:.
                  \label{gammaquerdeltan}
\ee

With the last relations
we obtain
the following approximate result for the averaged dimensionless conductance
in the case of maximal coupling strength ($\epsc = 0.5$) and no
Aharonov-Bohm flux penetrating the ring:
\bea
\la T_{0.5}^{\rm ana} \ra &=&
  \fr{ 16 \cos^2 \bar{\gamma} \sin^2 \Delta n \fr{\pi}{2} }
        {4 + \cos^4 \bar{\gamma} \li( 1 - \cos \Delta n \pi \re)^2}
        \label{avgtanap5} \\
  &=& \fr{ 16 } {1 + Q^2 }
    \fr{\sin^2 \li( \fr{\pi}{2} \sqrt{1+Q^2} \re)}
       {4 +
        \li( 1+Q^2 \re)^{-2} \li( 1 - \cos \li( \pi \sqrt{1+Q^2} \re) \re)^2 }
         \:. \nn
\eea

From this equation we immediately recover the zeros of
$\la T_{0.5}^{\rm ana} \ra$ at $\sqrt{1+Q^2}=2m$,
$m\in \mathbb{N}$. In Fig.~\ref{fig_avgT_p5}(b), Eq.~(\ref{avgtanap5}) is compared
 with the
numerical result, pointing out a good
agreement of the two curves, with deviations visible only close to the local maxima,
where  $\la T \ra$  is approximated by the
Lorentzian prefactor $1/(1+Q^2)$. The results for 1D rings presented in
Fig.~\ref{fig_avgT_p5} also properly describe a the 2D case with one
transverse channel. A comparison with the full numerical calculation (presented in
paper II of this series)
is given in the inset of Fig.~\ref{fig_3dinsetnoflux} and shows the very good agreement
between the 2D results and the 1D model with $\epsilon=0.316$.

Equation (\ref{avgtanap5}) holds in the strong coupling limit $\epsc = 0.5$.
For small coupling strengths, we find that
$\la T \ra$ is well described by the Lorentzian envelope function
\be\label{avgtenv}
\la T \ra_{\rm env} = 2 \,\fr{\epsc}{1-\epsc} \:\fr{1}{1+Q^2} \:,
\ee
as is demonstrated in Fig.~\ref{fig_avgT_p5}(a) for
$\epsc \le 0.4$.
The $\epsc$ dependence is identical to the case of spin-independent transport, see Eq.\ (\ref{avgtvonepsana}).
Deviations from this approximation now occur at the positions where $\la T \ra$ drops to zero,
and these dips become the narrower the smaller $\epsc$.

\begin{figure}[!t]
  \epsfxsize=8.5cm
  \centerline{\epsffile{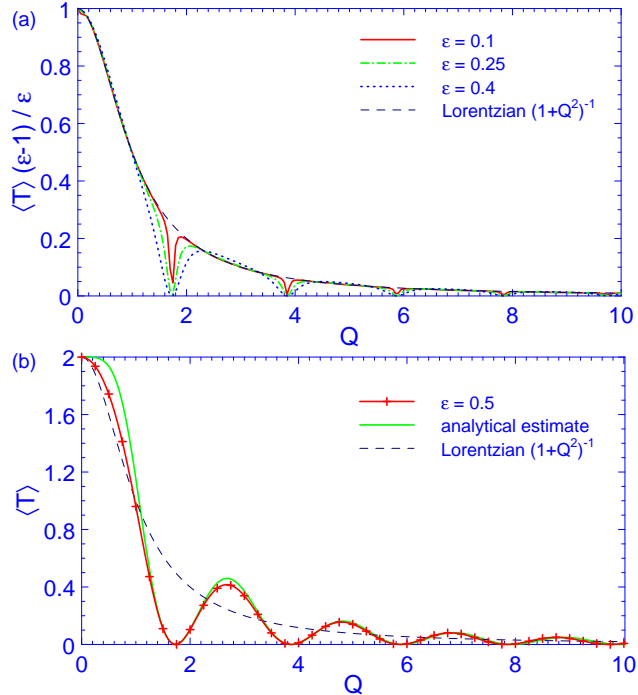}}
  \caption{
  (a) Energy-averaged dimensionless conductance $\la T\ra$ of Fig.~\ref{fig_3dinsetnoflux},
  scaled by a factor
  $(1-\epsc) / \epsc$
  for various coupling
  strengths $\epsc < 0.4$. Even though the precise shape around the minima depends
  strongly on the value of $\epsc$ (in particular, it becomes
  singular for $\epsc \to 0$), the overall decay away from the
  minima is very well  described by the Lorentzian envelope function (\ref{avgtenv}).
   (b)  Energy-averaged dimensionless conductance for maximal coupling strength
   $\epsc = 0.5$, in an in-plane (circular) magnetic field
   ($\varphi_t=\alpha=\pi/2$, $\phi^{\rm AB}=0$).
   The exact result (crosses) is well described by the analytical
   expression, Eq.~(\ref{avgtanap5}). For comparison, the Lorentzian
   decay is also shown.
  }
  \label{fig_avgT_p5}
\end{figure}

\begin{figure}[!t]
  \epsfxsize=8.5cm
  \centerline{\epsffile{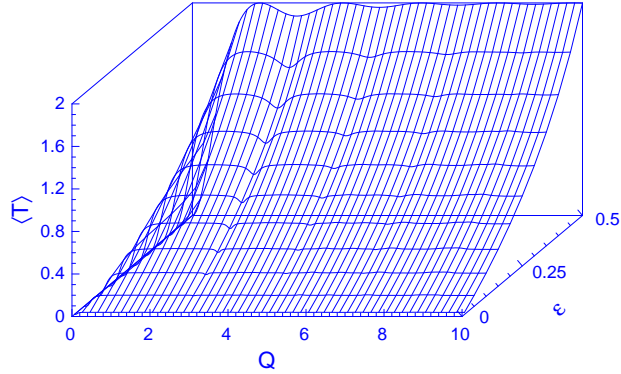}}
  \begin{center}
  \caption{
  Same as Fig.\ \ref{fig_3dinsetnoflux}, but
  for an Aharonov-Bohm flux $\phi^{\rm AB} = \phi^{\rm AB}_{\rm ext} =\phi_0/2$.
  For $Q=0$ the conductance is zero. As $Q \to \infty$, the
  oscillations diminish and $\la T \ra$ depends only on the
  coupling $\epsc$, saturating at  $\la T \ra = 2 \, \epsc / (1-\epsc) $.
  The minima of $\la T \ra$ occur at
  $Q=\sqrt{(2m+1)^2-1}$, $m\in \mathbb N$.}
  \label{avgT_3d_Bzp5}
    \end{center}
\end{figure}

So far, we investigated the situation where no Aharonov-Bohm flux was
penetrating the ring. The result in the presence of
a weak homogeneous magnetic
field $B_z$ corresponding to half a flux quantum is shown in
Fig.~\ref{avgT_3d_Bzp5}.
The averaged dimensionless conductance vanishes
in the diabatic limit $Q=0$.
This effect is the complement of the
asymptotic value $\la T\ra \to 0$ for $Q\to\infty$ in the case without
a magnetic flux, where
effectively half a flux quantum
can be provided as
geometric flux via the geometric phase, cf.~Eq.~(\ref{thetaabupdown}).
Note, however, that $\phi_0/2$
is the maximal possible geometric flux, achieved in the
adiabatic limit, and that geometric-phase effects are reduced in
non-adiabatic situations (see  Fig.~\ref{fig_3dinsetnoflux}).
In turn, an Aharonov-Bohm flux $\phi^{\rm AB}=\phi_0/2$ can be compensated
by the geometric flux in the adiabatic
limit, as is shown in Fig.~\ref{avgT_3d_Bzp5}.
 Furthermore, we now find a $\epsc / (1-\epsc)$-dependence of the
averaged dimensionless conductance
in the {\em adiabatic} limit $Q\to\infty$, which again is complementary to
the case $\phi^{\rm AB}=0$.
Also note that for $\phi^{\rm AB}=\phi_0/2$ local minima occur at the
positions $Q= \sqrt{(2m+1)^2 -1}$ with $m \in \mathbb N$, in contrast to
the minima at $Q=\sqrt{4 m^2 -1}$ for $\phi^{\rm AB}=0$.
However, for $\phi^{\rm AB}=\phi_0/2$ the dimensionless conductance at the
minima is only diminished and does not
drop to zero.

\subsection{Limiting situations: Adiabatic and diabatic regime}
\label{subsec_limiting}

Finally, we briefly investigate the dimensionless conductance in the
{\em adiabatic} and {\em diabatic}
limit for the in-plane magnetic field geometry. Adiabaticity is
characterized by the dominance of the magnetic field over the orbital motion,
manifesting itself in
\[
\gamma_\rho^\sigma= \pi/2
\]
(cf.~the geometric interpretation of the angles
$\gamma_\rho^\sigma$ in Fig.~\ref{fig_dreiecknprmub}).
Using $\zeta_\rho^\sigma = \gamma_\rho^\sigma/2$ in Eq.~(\ref{g1g2g3g4})
leads to the relation $g_1 = - g_4$ between the matrix elements
of the transfer matrices $\matt_1,\matt_{2}$.
When no additional Aharonov-Bohm flux is present, $\phi^{\rm AB} = 0$,
we obtain from Eq.~(\ref{deltas})
$\delta_3 = 0 $, yielding via Eq.~(\ref{a14a23noflux})
immediately
\[
T = 0
\]
for the dimensionless conductance  in the adiabatic limit
-- this is the result of the destructive interference
due to the maximal geometric phase.

For diabatic conditions, the magnetic field
does not play any role, and we are in the situation of
\[
\gamma_\rho^\sigma=0 \:,
\]
as again is clear from the geometric picture. This last equation still holds in
the case of a weak-to-moderate homogeneous magnetic field perpendicular to the
ring. After straightforward algebra, we recover an equation for the
dimensionless conductance  that is equivalent to the one derived by B\"uttiker
{\it et al.}~in Ref.~\onlinecite{imry} for a 1D ring subject to an external
Aharonov-Bohm flux,
\[
T_{\uarr \uarr}^{\rm diab}
= T_{\darr \darr}^{\rm diab}
= \fr{4 \epsc^2}{ 1 -
                       2 \cos \li( 2 \theta_d + 2 \theta_{\rm AB} \re)
                        (a+b)^2 + (a+b)^4  } \:,
\]
with the dynamic phase $\theta_d$ and the Aharonov-Bohm phase
$\theta_{\rm AB}$ introduced in Eq.~(\ref{thetadab}).

\section{Summary}
\label{sec_concl}

We have studied spin-dependent electronic transport in 1D rings in the presence
of inhomogeneous magnetic fields,
in particular for crown- and wire-like
magnetic fields with a varying homogeneous $B_z$ component
superposed.
We find characteristic signatures of geometric phases in the
magneto-conductance and in the energy-averaged transmissions.

In the adiabatic situation, the total conductance can be understood as
superposition of two independent electron gases
($||\suarr\ra\ra$ and $||\sdarr\ra\ra$)
that acquire different geometric (Berry)
phases due to the different orientation of their spin
with or against the magnetic field. This result can be understood within
a model of spin-independent transport plus spin-dependent Zeeman-interaction
and Berry phase,
cf.~Fig.~\ref{fig_mm_fulltransm}(a),(d).

In truly non-abiabatic situations,
the spins can flip, and the picture is more involved.
For the
magnetic field originating from a central micromagnet we have identified clear
signatures of geometric phases in the calculated magneto-conductance of 1D
rings,
see Fig.~\ref{fig_mm_fulltransm}(b)-(d). They appear as interference effects
that destroy the regular Aharonov-Bohm oscillations and become increasingly
visible towards the adiabatic limit.
Recent experiments \cite{ye} were still performed under
rather diabatic conditions that are not favorable for the observation of
geometric phases.

A generalization of the analytical approach presented here to
disorder effects and to diffusive one-dimensional rings 
appears as promising direction.

For the special case of an in-plane magnetic field
we investigated the
spin-flip effect of Refs.\ \onlinecite{2dlang,prl}, which originally
was uncovered in numerical simulations of 2D rings.
The 1D model allows for a rigorous analytical explanation and proof
of this quantum interference effect that
does not depend on the degree of adiabaticity.
For polarized incident electrons, a
small external Aharonov-Bohm flux can be
used to control spin flips and to tune
the polarization of transmitted electrons.
This spin-flip effect represents a promising
control tool in future spintronics engineering, since it also
prevails for rings with Rashba spin orbit coupling, \cite{FR03}
where the intrinsic effective magnetic field takes the role of
the applied inhomogeneous field in the present context.

\begin{acknowledgments}
We thank J.~Biberger, J.~Fabian and D.~Weiss for useful discussions.
MH gratefully acknowledges financial support from the Alexander von
Humboldt Foundation.
\end{acknowledgments}

\end{document}